\documentclass[a4paper,3p,11pt,sort&compress]{elsarticle}
\journal{Ultrasonics,  final version: 10.1016/j.ultras.2023.107112}
\date{}

\RequirePackage{snapshot}

\usepackage[Symbol]{upgreek}
\usepackage{mathptmx}
\usepackage[framed,numbered,autolinebreaks,useliterate]{mcode}
\usepackage[T1]{fontenc}
\usepackage[latin9]{inputenc}
\usepackage{amsmath}
\usepackage{amssymb}
\usepackage{graphicx}
\usepackage{setspace}
\usepackage{esint}
\PassOptionsToPackage{normalem}{ulem}
\usepackage{ulem}
\usepackage{marvosym}
\usepackage{multirow}
\usepackage{soul} 
\usepackage{physics}
\usepackage{xfrac}
\usepackage{dsfont}
\usepackage[hidelinks]{hyperref}
\usepackage{nomencl}
\usepackage{adjustbox}
\usepackage{mathtools}
\usepackage{nicefrac}
\usepackage[hang,flushmargin]{footmisc}

\soulregister\cite7
\soulregister\ref7
\soulregister\pageref7

\def\*#1{\mathbf{#1}}

\usepackage{hyperref}

\makeatletter

\newcommand{\vg}[1]{\bm{#1}}                        
\newcommand{\T}[0]{\mathrm{T}}                      
\newcommand{\Her}[0]{\mathrm{H}}                    
\newcommand{\I}[0]{\mathrm{i}}                      
\newcommand{\eps}[0]{\vg{\upepsilon}}               
\newcommand{\sig}[0]{\vg{\upsigma}}                 

\newcommand{\egv}[0]{\vg{\upphi}}                   
\newcommand{\Egv}[0]{\vg{\upPhi}}                   
\newcommand{\EgM}[0]{\vg{\Omega}}                   

\DeclareMathSymbol{\shortminus}{\mathbin}{AMSa}{"39}
\newcommand{\shortm}{\negthinspace\shortminus\negthinspace}

\newcommand{\eq}{\begin{eqnarray}}
\newcommand{\nq}{\end{eqnarray}}

\usepackage{subfig}
\usepackage{bm}
\usepackage{color}
\usepackage{epstopdf}

\graphicspath{{figures/}}

\newdefinition{rmk}{Remark}

\makeatother
\PassOptionsToPackage{numbers,sort&compress}{natbib}

\makenomenclature
\begin{document}

\newlength{\figwidth}
\setlength{\figwidth}{0.5\textwidth}
\newlength{\figheight}
\setlength{\figheight}{0.33\textwidth}

\title{Notes on osculations and mode tracing in semi-analytical\\ waveguide modeling}

\author[cimne]{Hauke~Gravenkamp\corref{cor1}}
\ead{hgravenkamp@cimne.upc.edu}

\author[lj]{Bor~Plestenjak}

\author[pa]{Daniel~A.~Kiefer}

\address[cimne]{International Centre for Numerical Methods in Engineering (CIMNE), 08034 Barcelona, Spain}

\address[lj]{Faculty of Mathematics and Physics, University of Ljubljana, Jadranska 19, SI-1000 Ljubljana, Slovenia}

\address[pa]{Institut Langevin, ESPCI Paris, Universit\'e PSL, CNRS, 75005 Paris, France}

\cortext[cor1]{Corresponding author}

\begin{abstract}\noindent
    The dispersion curves of (elastic) waveguides frequently exhibit crossings and osculations (also known as veering, repulsion, or avoided crossing). Osculations are regions in the dispersion diagram where curves approach each other arbitrarily closely without ever crossing before veering apart. In semi-analytical (undamped) waveguide models, dispersion curves are obtained as solutions to discretized parameterized Hermitian eigenvalue problems. In the mathematical literature, it is known that such eigencurves can exhibit crossing points only if the corresponding matrix flow (parameter-dependent matrix) is uniformly decomposable. We discuss the implications for the solution of the waveguide problem. In particular, we make use of a simple algorithm recently suggested in the literature for decomposing matrix flows. We also employ a method for mode tracing based on approximating the eigenvalue problem for individual modes by an ordinary differential equation that can be solved by standard procedures.
\end{abstract}
\maketitle

\section{Overview}
\begin{figure}[bt]\centering
    \subfloat[\label{fig:dispersionSlightlyInhomogeneous_a}]{ \includegraphics[height = 0.255\textwidth]{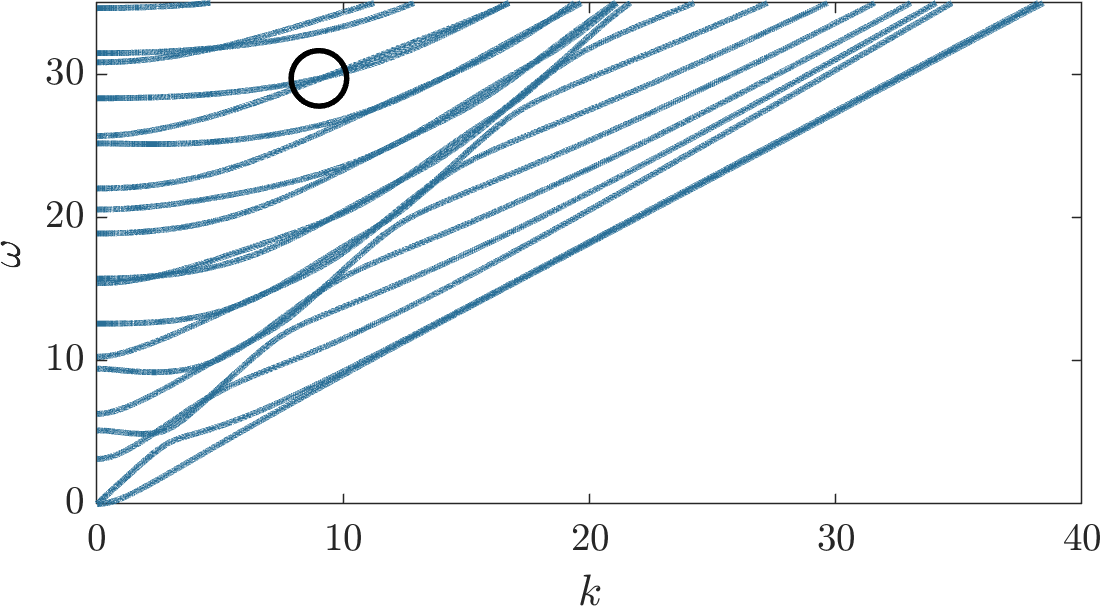}}\hfill
    \subfloat[\label{fig:dispersionSlightlyInhomogeneous_b}]{ \includegraphics[height = 0.255\textwidth]{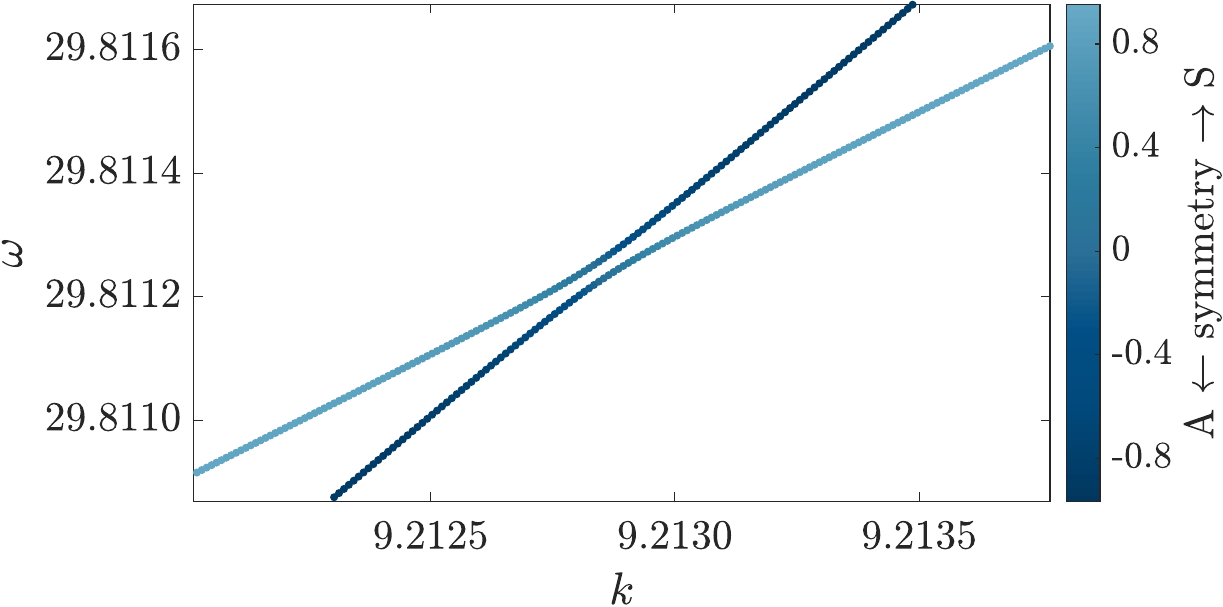}}
    \caption{Dispersion diagram of Lamb waves in a slightly inhomogeneous plate and magnification of an osculation (in the region indicated by the circle in (a)). One half of the plate has a 1\% larger shear modulus than the other. \label{fig:dispersionSlightlyInhomogeneous}}
\end{figure}

\noindent
It is well known among researchers working on the simulation or experimental application of guided waves that the dispersion curves of such systems may cross or, in other cases, approach each other very closely without crossing. The latter situation is often referred to as \textit{osculation, mode veering, mode repulsion, avoided crossing, eigenvalue avoidance}, or even \textit{mode kissing}. These two cases can be difficult to distinguish in the dispersion diagram since what looks like a crossing at first sight, may reveal to be an osculation only after significant magnification; see Fig.~\ref{fig:dispersionSlightlyInhomogeneous} for an example. In fact, we will see that the distance between two seemingly crossing modes can, in principle, be arbitrarily small and may only be limited by the machine precision when putting numerical methods into practice. 

When considering the eigenvectors (representing the mode shapes of waves propagating along the guide), the differences become even more subtle as, oftentimes, the eigenvector of the one mode just before the osculation is rather similar to that of the other mode after the osculation \cite{Mace2012}. This phenomenon is visualized in Fig.~\ref{fig:dispersionSlightlyInhomogeneous_b}, where two modes rapidly switch their property of being almost symmetric/antisymmetric with respect to a waveguide's symmetry line.\footnote{We decompose the normalized eigenfunctions into their even and odd parts. The shown ``level of symmetry'' is the difference between the square integral of the even part and the odd part; for details, see \ref{sec:symmetry_of_guided_waves}.} It can seem like a futile task to try and connect the discrete points in a dispersion diagram in a way that consistently represents individual modes and to determine precisely whether a point represents a crossing or an osculation. 
We may even ask \textit{'Does it matter?'} -- ultimately, any experiment will involve a finite frequency and time interval, and a numerical simulation will usually include all resolvable modes at a given frequency, irrespective of their relationship to modes at other frequencies. Nevertheless, there is a desire to shed some light on this issue, even if it were just for the seemingly mundane purpose of establishing a consistent naming convention for the individual modes. Beyond that, researchers have been fascinated with such mode-veering or osculation phenomena, as indicated by countless essays on the occurrence, relevance, and interpretation of these effects.\footnote{We should note that a different definition of an \textit{osculating curve} exists in the scope of differential geometry where it relates to curves that do contact one another -- in contrast to the phenomenon of osculations in the context of dispersion curves. } 

A particularly basic -- yet not any less confusing -- example may help to illustrate the problem: Consider an unbounded homogeneous linearly elastic isotropic plate with traction-free surfaces (Lamb's problem). It is well known that the Lamb wave modes can be separated into two sets of symmetric and antisymmetric ones (with respect to the midplane). Numerically, we can compute these sets individually by discretizing only one half of the plate's thickness and applying symmetric/antisymmetric boundary conditions -- i.e., fixing the displacement first in the $x$-direction and then in the $y$-direction at the symmetry line. Plotting all modes in one graph reveals that many of the dispersion curves belonging to different sets cross each other. As we can compute the two sets individually, we can be certain that these are indeed crossing points rather than osculations. This approach, of course, is only applicable if the structure of interest has such symmetries. 
We may, for instance, consider a similar plate, with the only difference being that the upper and lower half have slightly different material parameters. Hence, we have no other option than to model the full plate as one. In this case, it is known that the former crossing points now become osculations -- one such osculation for exactly this problem is shown in Fig.~\ref{fig:dispersionSlightlyInhomogeneous_b}. This behavior may seem unintuitive: the difference in material parameters can be arbitrarily small; still, in theory, we obtain
crossing points only if this difference is precisely zero. In a discretized numerical system, we could argue that it is mainly a matter of accuracy whether or not we are able to resolve such an osculation. For details on this and other examples, we refer to the work by Kausel et al.~\cite{Kausel2015}, who meticulously derived the conditions for the existence of osculations (or absence of double-roots) for several systems, including the aforementioned homogeneous plate. The case of homogeneous isotropic plates is also discussed in the earlier papers by Veres et al.~\cite{veres_crossing_2014} as well as Zhu\,\&\, Mayer~\cite{zhu_crossing_1993}. Discussions on how to predict the mode segregation based on material symmetries in the case of plates can also be found in \cite{kiefer_elastodynamic_2022,hernando_quintanilla_symmetry_2017}.  

\noindent 
In this current communication, we want to look at this topic from a different angle and apply results that have been discovered by mathematicians in the context of the analysis of certain matrix functions or \textit{matrix flows}. Though the terminology is not unique, we use \textit{matrix flows} as in the papers by Uhlig et al.\  \cite{Uhlig2020,Uhlig2019} to refer to matrix functions that depend continuously on one parameter, say $\*E(k)$. Alternatively, such a function has simply been called \textit{parameter-dependent matrix}. Often, the parameter is associated with a time variable, while, in our case, we will identify it as a wavenumber, denoted as $k$. The matrix $\*E(k)$ can then be interpreted as a wavenumber-dependent \textit{stiffness matrix} at a given location along the waveguide. Since the works of von Neumann and Wigner \cite{Neumann1930}, it is known that the eigencurves of the type of matrix flow considered here (to be defined later) cannot cross unless it is \textit{uniformly decomposable}. Uhlig \cite{Uhlig2020} points out that this crucial condition is frequently overlooked, leading to significant confusion on this topic. 
Decomposable means that a matrix can be block-diagonalized by a similarity transformation. In the case of a matrix flow, we are interested in whether it can be \textit{uniformly} decomposed, i.e., whether there is a transformation that leads to proper block-diagonal matrices with the same block structure for any value of the parameter $k$. \textit{Proper block-diagonal} means that the matrix consists of more than one block. If this is the case, we can compute the eigencurves of each block separately. For each (indecomposable) block, we then know that the eigencurves do not cross, making the sorting and labeling of eigenvalues trivial. As we have already seen, the waveguide problem may or may not exhibit eigencurve crossing, depending on the geometry and material. Thus, the corresponding matrix flows should reveal such a structure. 

These theoretical results, albeit extremely useful, appear to have remained unnoticed in the field of waveguide modeling. In fact, many authors proposed methods for \textit{mode tracing} in the post-processing, i.e., for following the individual dispersion curves to determine which of the discrete computed points in the frequency-wavenumber-plane belong to the same mode. A typical idea is to employ the modal assurance criterion (MAC); see, e.g., \cite{Allemang2003} for an overview. This approach establishes a measure for the `similarity' of eigenvectors and assumes that two mode shapes with a large MAC value belong to the same mode (which may be misleading in the vicinity of osculations as mentioned above). Other authors have made use of extrapolation using Taylor- or Pad\'e-expansion \cite{Gravenkamp2013a}, where derivatives are either approximated by finite differences from previous points in the dispersion diagram or by evaluating eigenvalue derivatives directly~\cite{Krome2016}.
Applying the concept of block-diagonalization, we can eliminate the need for such post-processing efforts if we can determine -- at least numerically for a given matrix flow --  the decomposability of the involved matrices and compute the corresponding transformation matrices. For this purpose, we will follow a pragmatic approach recently suggested by Uhlig \cite{Uhlig2022} for general matrix flows. Uhlig computes the eigenvalue decomposition of the matrix at some value of the parameter, say $\*E(k_\mathrm{a})$, and applies the eigenvectors to transform the matrix flow at some other value $\*E(k_\mathrm{b})$. By re-sorting the rows and columns, it is easy to assess whether the resulting transformed matrix can be written in block form. If so, we compute the eigenvalues of each separate block. The idea is that, as long as we are successful in applying the same transformation based on the same eigenvectors computed at $k_\mathrm{a}$ to obtain the same block structure, we know that the eigencurves that belong to the same block do not cross, and further post-processing is unnecessary. This is certainly true when all eigenvalues of $\*E(k_\mathrm{a})$ are distinct; the special case of repeated eigenvalues is discussed in Section 3.3.

We emphasize that the discussion in this paper is, for now, limited to Hermitian matrix flows, which occur in the common semi-analytical models of acoustic and elastic undamped waveguides. By discretizing the waveguide's cross-section, these approaches are quite generally applicable to complex geometries and inhomogeneous materials. However, we exclude damping (be it through viscoelastic materials or radiation into an unbounded domain) \cite{Bartoli2006,Mazzotti2013b,Gravenkamp2014b,Gravenkamp2015,Gravenkamp2014c}, which leads to non-Hermitian systems.

Since we are touching on the topic of mode tracing, we also include in Section~\ref{sec:trace} the application of another very interesting approach to the computation of eigencurves. This approach came, once again, to our attention through the works of Uhlig on matrix flows \cite{Uhlig2019} and, once again, appears to be unknown in the context of waveguide modeling. The idea is to rewrite the eigenvalue problem (for one particular mode) as an ordinary first-order matrix differential equation in the parameter $k$. This differential equation can be easily solved by standard algorithms, using known solutions at some $k$ as initial values. 
We are not yet sure about this approach's practical usefulness in the context of the waveguide problem. It can be computationally efficient in certain cases where a limited number of modes is of interest. In any case, we believe that this approach will be interesting to researchers working on guided waves, as it gives a very different perspective on the solution of the dispersion curve problem.

\section{Problem statement}\noindent
We consider wave propagation in linear elastic media along a structure of constant cross-section. Hence, the governing equation formulated for the stress tensor $\sig$, in the absence of body forces, is 
\begin{equation}
    \nabla \cdot \sig - \rho \ddot{\*u}= \*0 
\end{equation}
with the assumed linear material law being defined by the constitutive tensor $\*C$
\begin{equation}
    \sig = \*C : \eps \ ,
\end{equation}
and the strains are obtained by the standard linear relation from the displacement vector $\*u$
\begin{equation}
    \eps = \frac{1}{2}(\nabla \*u + \nabla \*u^\T) \  .
\end{equation}
As usual, $\rho$ denotes the mass density, $\ddot{\*u}$ is the acceleration, $\nabla$ is the Nabla operator, and $(\cdot)^\T$ indicates the transpose. Note that the governing equations hold in two and three dimensions with the displacement vector consisting of the entries $(u_x(x,y,t),u_y(x,y,t))$ or $(u_x(x,y,z,t),u_y(x,y,z,t),u_z(x,y,z,t))$, respectively, and the corresponding second-order stress tensor of dimension $2\times2$ or $3\times3$. 
We use a semi-analytical method to compute the modes that can propagate along the waveguide. That is to say, we employ a weak form of the governing equation only on the waveguide's cross-section and discretize the same in the finite element sense. Along the propagation direction, the problem remains continuous. While there are countless methods to compute dispersion curves of elastic waveguides (e.g., \cite{Mead1973b,Renno2013b}), we are interested in those that lead to a matrix function of the wavenumber, involving (for now) only constant coefficient matrices. In addition, we wish to restrict the discussion to Hermitian eigenvalue problems.
There are several closely related methods to achieve this goal, in particular, the thin layer method (TLM) \cite{Kausel1981a,Barbosa2012a,Kausel2004}, scaled boundary finite element method (SBFEM) \cite{Gravenkamp2012,Gravenkamp2014f,Itner2020a} and semi-analytical finite element method (SAFE) \cite{Hayashi2006b,Hayashi2003a,Thakare2017}. For the simple application discussed here where only systems of a constant cross-section in the frequency domain are addressed, these methods can be considered to be equivalent.\footnote{In a nutshell, the TLM was developed in the context of simulating soil layers \cite{Kausel1981a}. The SBFEM has its origin in the TLM \cite{Wolf1994} and generalizes it to more complex (polytopal) domains \cite{Song1997,Gravenkamp2018a}. The SAFE method was developed independently in the context of ultrasound simulation but uses the same concept as the TLM \cite{Gavric1994,Bartoli2006}. Interesting differences exist in the application of these methods to more complex geometries and materials, but we will not delve into the details here.} For computing the modes, we formally perform a two-dimensional Fourier transform such that we obtain a formulation in the $(\omega, k)$-domain, where $\omega$ is the angular frequency, and $k$ denotes the wavenumber in the direction of wave propagation along the waveguide. 
We refrain from presenting the details of the derivation as it can easily be found in the literature on the aforementioned methods. 
Here, we are mainly interested in the resulting eigenvalue problem and the properties of the matrices involved.
After the discretization and transformation outlined above, we obtain an eigenvalue problem depending on the parameters $\omega$ and $k$ in the form
\begin{equation}\label{eq:quadEVP}
   (k^2 \*E_0 - k \*E_1 + \*E_2) \egv = \omega^2\*M \egv\ .
\end{equation}
Here, the eigenvector $\egv$ contains the coefficients (typically nodal displacements) of the finite-element discretization on the cross-section\footnote{Various different interpolants can be employed as trial and test functions, such as the standard Lagrange interpolation polynomials, hierarchical shape functions, NURBS and other splines, or moving least squares; see \cite{Gravenkamp2019} for an overview in the context of the SBFEM. While the approximations have certain advantages and drawbacks, these differences are not relevant to the current discussion. Here, we can, for all means and purposes, assume that the cross-section is discretized by basic node-based shape functions; hence, the eigenvector contains nodal displacements.} for a given mode, and the matrices $\*E_0,\, \*E_1,\, \*E_2,\, \*M$ are finite-element matrices computed by numerical integration over the cross-section. From a finite element point of view, $\*M$ is nothing but the mass matrix of the cross-section, while $(k^2 \*E_0 - k \*E_1 + \*E_2)$ represents a wavenumber-dependent stiffness matrix.
The following properties are observed (in the absence of material damping): 
\begin{itemize}  \setlength\itemsep{-0.3\baselineskip}
    \item $\*E_0$ and $\*M$ are positive definite, $\*E_2$ is positive semi-definite.
    \item $\*E_1$ is imaginary, while $\*E_0$, $\*E_2$, $\*M$ are real.  
    \item All four matrices are Hermitian (i.e., $\*E_i^\Her=\*E_i$, $\*M^\Her=\*M$).
\end{itemize}
We use $(\cdot)^\Her$ to denote the Hermitian (or conjugate) transpose. We highlight again that we do not consider material damping in the current work, as its inclusion is typically done by introducing complex-valued material parameters \cite{Manconi2013b}, rendering the eigenvalue problem non-Hermitian. This, in turn, has strong implications for the following approaches to matrix decomposability.
\begin{rmk}
    This eigenvalue problem is often modified by writing it for $\I k$ instead of $k$, making all matrices real-valued. This substitution can be useful, particularly when computing the wavenumbers at a given frequency. However, in that case, we have a term involving $\I\,\*E_1$, which is not Hermitian. Also, since $\*M$ is invertible, it may seem convenient to multiply Eq.~\eqref{eq:quadEVP} by $\*M^{-1}$, but this again may lead to a non-Hermitian system. As Hermitian matrices are crucial for the following algorithms to be applied effectively, we will address the generalized eigenvalue problem in the above form. Let us note, however, that since $\*M$ is positive definite, we could 
    transform Eq.~\eqref{eq:quadEVP} into the equivalent Hermitian system
    \begin{equation}\label{eq:quadEVP_B}
   (k^2 {\*{\widehat E}_0} - k {\*{\widehat E}_1} + {\*{\widehat E}_2})\, \widehat\egv = \omega^2\, \widehat\egv\ ,
\end{equation}
where $\widehat \egv = \*M^{\nicefrac{1}{2}}\egv$ and ${\*{\widehat E}_i}=\*M^{\shortminus{\nicefrac{1}{2}}}\,\*E_i\,\*M^{\shortminus{\nicefrac{1}{2}}}$ for $i=0,1,2$. In contrast to \eqref{eq:quadEVP}, when computing $\omega^2$ for a given $k$, the above formulation is a standard eigenvalue problem.
\end{rmk}
\begin{rmk}
    In addition to the semi-analytical methods listed above, another popular one is the spectral collocation method \cite{Adamou2004,Kiefer2019,kiefer_gew_2022}. While this method leads to an eigenvalue problem of the same form as Eq.~\eqref{eq:quadEVP}, the properties of the matrices are different. Perhaps most importantly, it usually leads to a singular system. Hence, we exclude this method from the current discussion. Furthermore, an alternative for analyzing the modal behavior consists in discretizing a section of the waveguide rather than only its cross-section. This approach, sometimes referred to as Wave Finite Element Method, is beneficial for structures that are periodic (but not homogeneous) \cite{Droz2014,Mitrou2016,Zhu2023}. However, it leads to a matrix structure different from the one we address in this work. 
\end{rmk}
The two-parameter eigenvalue problem \eqref{eq:quadEVP} has infinitely many solutions, i.e., there exist infinitely many combinations of frequencies and wavenumbers for which the eigenvalue problem is satisfied. The task is usually to compute the wavenumbers for a given frequency range or vice versa (depending on the application, one of the two approaches may be preferred). To compute the dispersion curves, we are interested in the real-valued solutions of the wavenumber (in the absence of damping). In this case, it is usually more convenient to define a range of wavenumbers and compute the corresponding frequencies, as Eq.~\eqref{eq:quadEVP} can be treated as a linear eigenvalue problem for $\omega^2$, while it is quadratic in $k$.
We may then abbreviate the eigenvalue problem as
\begin{equation}\label{eq:evpFlow}
    \*E(k) \egv  = \bar{\omega}\*M \egv
\end{equation}
with $\bar{\omega} = \omega^2$ and $\*E(k) =k^2 \*E_0 - k \*E_1 + \*E_2$, establishing a generalized eigenvalue problem of the matrix flow $\*E(k)$. Alternatively, we will write
\begin{equation}\label{eq:evpFlowM}
    \*E(k) \Egv  = \*M \Egv \EgM
\end{equation}
with the matrix $\Egv$ containing all eigenvectors as columns and the diagonal matrix $\EgM$, which contains the corresponding eigenvalues on the diagonal. For ease of notation, we explicitly write only $\*E(k)$ as a function of $k$ while the resulting eigenvectors and eigenvalues, of course, depend on $k$ as well. Note that $\*M$ is a constant matrix.

\section{Computing dispersion curves utilizing block-diagonal decompositions}\label{sec:decompose}

\subsection{Decomposability of generalized Hermitian matrix flows }\noindent
In order to apply the method of decomposing the matrix flow and solving each block separately, we make a straightforward extension of the approach in \cite{Uhlig2022} to generalized eigenvalue problems.
Consider the matrix flow, Eq.~\eqref{eq:evpFlowM}, at a given value of $k$, say $k_\mathrm{a}$, and compute the corresponding eigenvector and eigenvalue matrices as
\begin{equation}
    \*E(k_\mathrm{a}) \Egv_\mathrm{a}  = \*M \Egv_\mathrm{a} \EgM_\mathrm{a} \ .
\end{equation}
Since $\*E(k_\mathrm{a})$ is Hermitian, and $\*M$ is Hermitian positive definite, both $\Egv_\mathrm{a}^\Her \*E(k_\mathrm{a}) \Egv_\mathrm{a}$ and $\Egv_\mathrm{a}^\Her \*M \Egv_\mathrm{a}$ are diagonal, and it is common to normalize the eigenvectors such that
    \begin{subequations}\label{eq:normalization}
        \begin{align}
        \Egv_\mathrm{a}^\Her \*E(k_\mathrm{a}) \Egv_\mathrm{a} &= \EgM_\mathrm{a}\ , \label{eq:normalization_a}\\
        \Egv_\mathrm{a}^\Her \*M      \Egv_\mathrm{a} &= \*I \ . \label{eq:normalization_b}
    \end{align}
    \end{subequations}
Hence, the matrices $\*E(k)$ and $\*M$ are diagonalizable for any $k$ by a similarity transformation using the corresponding eigenvectors (which depend on $k$). On the other hand, applying the same transformation defined by those eigenvectors obtained at $k_\mathrm{a}$ to the matrix flow at a \emph{different} value $k_\mathrm{b}$, i.e., 
    \begin{align}
    \Egv_\mathrm{a}^\Her \*E(k_\mathrm{b}) \Egv_\mathrm{a} &= \*B(k_\mathrm{b}) \ ,\label{eq:normalization2_a}
\end{align}
we obtain some matrix $\*B(k)$ that will generally not be diagonal (in contrast, Eq.~\eqref{eq:normalization_b} obviously holds for any $k$ as $\*M$ is independent of $k$). 
The matrix flow is uniformly decomposable if there exists a fixed similarity transformation such that the resulting $\*B(k)$ is a proper block-diagonal matrix (after appropriate rearrangement) with the same block structure for all $k$.
The algorithm suggested by Uhlig applies the transformation~\eqref{eq:normalization2_a} based on the eigenvector matrix $\Egv_\mathrm{a}$ to the matrix flow at a different value $k_\mathrm{b}$ within the interval of interest. It then block-diagonalizes the resulting matrix $\*B(k_\mathrm{b})$ by re-arranging its rows and columns.
Due to numerical errors in the computation of the finite element matrices and the eigenvectors, we should define a threshold below which small values in $\*B(k)$ are treated as zeros in order to reveal the block structure.
If the resulting matrix consists of more than one block on the diagonal, we identify the columns of $\Egv_\mathrm{a}$ corresponding to each block 
\begin{equation}
    \Egv_\mathrm{a} = \{\Egv_\mathrm{a,1},\Egv_\mathrm{a,2},...,\Egv_\mathrm{a,m}\}
\end{equation}
where $m$ is the number of blocks, and $\Egv_\mathrm{a,i}$ are matrices formed by $n_i$ eigenvectors with $n_i$ being the size of each block.
When computing the eigencurves (in our case, dispersion curves) of the matrix flow, we do so for the individual blocks separately, i.e.,
\begin{equation}
    \*{\bar{E}}_i(k) \vg{\upPsi}_i  = \*{\bar{M}}_i \vg{\upPsi}_i \EgM_i
\end{equation}
with 
\begin{subequations}
    \begin{align}
        \*{\bar{E}}_i(k) &= \Egv_{\mathrm{a},i}^\Her \*E(k) \Egv_{\mathrm{a},i}\ , \\
        \*{\bar{M}}_i &= \Egv_{\mathrm{a},i}^\Her \*M \Egv_{\mathrm{a},i}\ .
    \end{align}
\end{subequations}

\subsection{Minimal example}\label{minEx}
\begin{figure}\centering
    \subfloat[\label{fig:minExDiri}]{ \includegraphics[width = 0.47\textwidth]{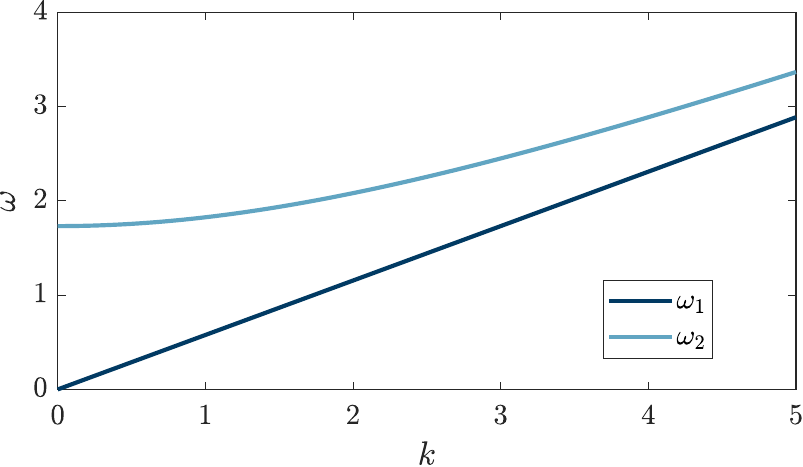}}\hfill
    \subfloat[\label{fig:minExFree}]{ \includegraphics[width = 0.47\textwidth]{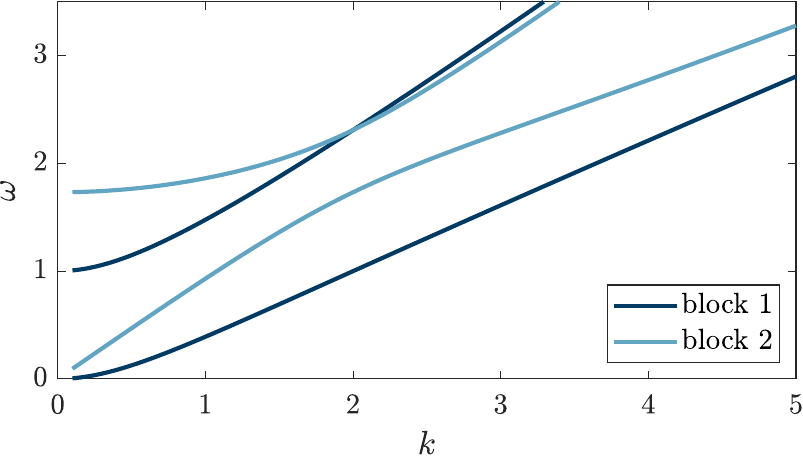}}
    \caption{Dispersion curves of a homogeneous isotropic plate, discretized by one linear element: (a) fixed horizontal displacements, (b) free surfaces.}
\end{figure}

\noindent
Consider a homogeneous isotropic plate of thickness $h=2$ in a plane strain approximation and assume that the horizontal displacements are fixed on both the upper and lower surface. As a first very rough approximation, we discretize the thickness direction by only one linear finite element, thus interpolating the displacement components by the two shape functions 
\begin{equation}
    N_1(y) = \frac{1-y}{2}, \quad N_2(y) = \frac{1+y}{2}.
\end{equation}
After applying the Dirichlet boundary conditions, the coefficient matrices are of dimension $2\times 2$ and can be calculated analytically (see, e.g., \cite{Kausel1977}). For convenience, we choose the material parameters: shear modulus $G=1$, mass density $\rho=3$, Poisson's ratio $\nu=0.25$, leading to the simple expressions
\begin{equation}
    \*M = \begin{bmatrix}
        2 & 1 \\ 
        1 & 2 
    \end{bmatrix}, \quad
    \*E_0 = \frac{1}{3}\begin{bmatrix*}
        2 & 1 \\ 
        1 & 2 
    \end{bmatrix*},\ \quad 
    \*E_1 = \*0, \quad 
    \*E_2 = \frac{3}{2}
    \begin{bmatrix*}[r]
        1 & \shortm1 \\ 
        \shortm1 & 1 
    \end{bmatrix*}\ .
\end{equation}
It is easy to verify that the eigenvectors can be written as (normalized to satisfy Eqs.~\eqref{eq:normalization})
\begin{equation}
    \egv_1 =\frac{1}{\sqrt{6}}\begin{bmatrix}
        1  \\ 
        1  
    \end{bmatrix}, \quad 
    \egv_2 =\frac{1}{\sqrt{2}}
    \begin{bmatrix*}[r]
        1  \\ 
        \shortm1  
    \end{bmatrix*}
\end{equation}
with the corresponding eigenvalues
\begin{equation}\label{eq:minEx_curves}
    \omega^2_1 = k^2/3, \quad \omega^2_2 = k^2/3 + 3\ .
\end{equation}
The resulting dispersion curves are plotted in Fig.~\ref{fig:minExDiri}. 
As the eigenvectors are independent of $k$, they decompose (in this case, even diagonalize) the matrix flow for any $k$. Hence, we obtain two decoupled problems for the two modes:
\begin{equation}\label{eq:minExNorm}
    \Egv^\Her \*E(k) \Egv = \begin{bmatrix}
        k^2/3 & 0 \\ 
        0 & k^2/3 + 3 
    \end{bmatrix} \ .
\end{equation}
Note that while the matrix flow describing the Lamb wave problem (homogeneous isotropic plate) is generally decomposable into two blocks, the property of constant eigenvectors here is a consequence of the simple linear interpolation and will generally not hold when using a more accurate discretization. The decomposability, in turn, implies the possibility of mode crossing. However, in this particular example, the curves defined by Eq.~\eqref{eq:minEx_curves} obviously do not cross. 
If, on the other hand, we consider the complete $4\times 4$ matrices, i.e., without fixing the horizontal displacements at the surfaces, the resulting matrix flow can be decomposed into two $2\times 2$ blocks and exhibits crossing modes, see Fig.~\ref{fig:minExFree}. In this case, the analytical calculation is already laborious and leads to lengthy expressions. Hence, we present only an example of numerical results of the decomposition for the two values $k_\mathrm{a}=1$ and  $k_\mathrm{b}=2$:
\begin{equation}\label{eq:minExFreeDecomp}
    \Egv^\Her \*E \Egv\Big|_{k=1} \approx \begin{bmatrix}
        0.15 & 0    &  0    & 0    \\
        0    & 0.86 &  0    & 0    \\
        0    & 0    &  2.18 & 0    \\
        0    & 0    &  0    & 3.47
    \end{bmatrix} \,, \qquad
    \Egv^\Her \*E \Egv\Big|_{k=2} \approx \begin{bmatrix}
        1    & -0.09 & 0    & 0      \\
       -0.09 &  \phantom{+}5.33 & 0    & 0      \\
        0    &  0    & 3.51 & 0.96   \\
        0    &  0    & 0.96 & 4.83
    \end{bmatrix} \,.
\end{equation}
As a remark, let us appreciate the interesting case of the so-called shear-horizontal (SH or out-of-plane) modes in a homogeneous plate. Their eigenvectors are independent of $k$, even for arbitrarily fine discretization. Hence, the matrix flow can be diagonalized, and the eigencurves can be computed separately for each mode based on the solution at any given value of $k$ as discussed in detail in \cite{Gravenkamp2016a}. However, such SH modes are known to not exhibit any crossings or osculations~\cite{Kausel2015}. Hence, shear-horizontal modes serve as an extreme case in two ways: Firstly, their matrix flow can be exactly diagonalized, which is the finest possible decomposition in this sense. Secondly, judging only from the decomposability criterion, one would conclude that  \emph{all of their dispersion curves are allowed to cross} each other. However, it turns out that \emph{none of them cross}.

\subsection{Repeated eigenvalues}\noindent
While the simple algorithm for testing the decomposability of a matrix flow, as summarized above, works remarkably well for most cases, it has one pitfall that appears to have been overlooked in the previous literature: If any of the eigenvalues in $\EgM_\mathrm{a}$ computed at the arbitrarily chosen value $k_\mathrm{a}$ are not distinct, it can happen that the computed eigenvectors $\Egv_\mathrm{a}$ do not block-diagonalize the matrix flow at values other than $k_\mathrm{a}$, even though there is some other similarity transformation that does. This is because the eigenvectors corresponding to the repeated eigenvalues span a subspace, with any linear combination being again an eigenvector. Only using the solution at $k_\mathrm{a}$, it is then generally not possible to find the eigenvectors in the subspace that block-diagonalize the matrix flow for other values of $k$. 
Note that the existence of repeated eigenvalues itself reveals information about the decomposability. Repeated eigenvalues occur either where two modes cross at isolated points (indicating decomposability) or in special cases where curves are identical. 
We will see a practical example of the latter case in Section~\ref{numEx_3D}.
Let us first clarify the problem related to repeated eigenvalues with the help of another minimal example. As a small modification of the example discussed in Section~\ref{minEx}, consider the matrix flow
\begin{equation}\label{eq:repEVflow}
\*E(k) = \frac{k^2}{4}
\begin{bmatrix}
    3 & 2 \\
    2 & 3
\end{bmatrix}    +
\frac{1}{2}
    \begin{bmatrix*}[r]
        1  & \shortm1\\ 
        \shortm1  & 1
    \end{bmatrix*}
\end{equation}
with the eigenvalues and eigenvectors
\begin{equation}
    \EgM(k) = \begin{bmatrix}
        5k^2/4 & 0 \\ 
        0 & k^2/4 + 1 
    \end{bmatrix}, \quad
    \Egv =\frac{1}{\sqrt{2}}
    \begin{bmatrix*}[r]
        1  & \shortm1\\ 
        1  & 1
    \end{bmatrix*}.
\end{equation}
\begin{figure}\centering
    \includegraphics[width=0.47\textwidth]{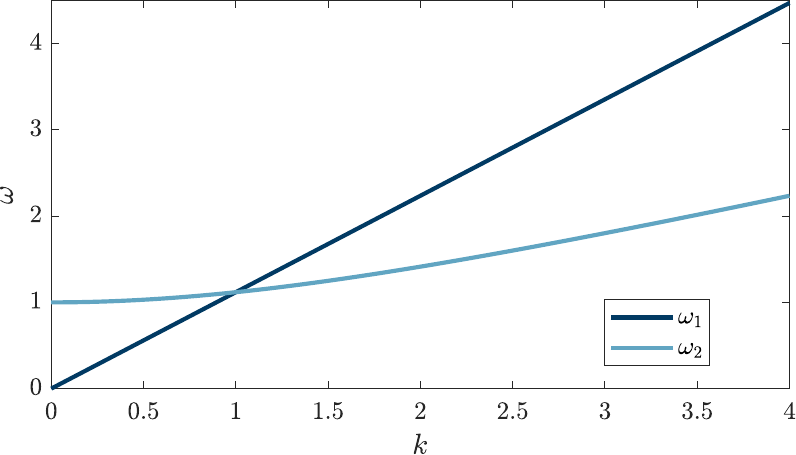}
    \caption{Eigencurves of the matrix flow defined by Eq.~\eqref{eq:repEVflow}. \label{fig:repEV}}
\end{figure}
The two eigencurves cross at $k=1$ (see Fig.~\ref{fig:repEV}), since
\begin{equation}
    \EgM(1) = \begin{bmatrix}
        5/4 & 0 \\ 
        0 & 5/4 
    \end{bmatrix} \ .
\end{equation}
At the crossing point, we also obtain
\begin{equation}
    \*E(1)=\EgM(1) = \begin{bmatrix}
        5/4 & 0 \\ 
        0 & 5/4 
    \end{bmatrix} \ .
\end{equation}
Since the eigenvalues at $k=1$ are identical, any linear combination of the eigenvectors $\Egv$ is an eigenvector of the matrix flow. In particular, if we computed the eigenvectors at $k=1$, an obvious choice would be (as $\*E(1)$ is diagonal)
\begin{equation}
\Egv(1) =\begin{bmatrix}
    1  & 0\\ 
    0  & 1
\end{bmatrix}
\end{equation}
which clearly does not diagonalize the matrix flow $\*E(k)$ at any value other than $k=\pm 1$.
We shall note that, to our practical application of identifying mode crossing and osculations in the computation of dispersion curves, such repeated eigenvalues do not pose a significant difficulty. Recall that we test the decomposability of the matrix flow using the complete solution of the eigenvalue problem \eqref{eq:evpFlowM} at some value $k_\mathrm{a}$. If this solution exhibits repeated (i.e., identical) eigenvalues, they indicate that their eigencurves cross at this point (unless the curves are identical everywhere, in which case we may choose not to refer to those curves as crossing). Consequently, we know that they belong to a block that may be further decomposed. When proceeding with the computation of the eigencurves for each block, we can continue testing for the decomposability of each block at consecutive values of $k$. If the repeated eigenvalues are due to mode-crossing, we can expect the values to differ already at the following $k$-values, and, hence, we will usually find a suitable decomposition in the next step. If, on the other hand, we obtain the same number of repeated eigenvalues over the entire range of $k$-values, we can assume that these belong to repeated eigen\textit{curves}. These are identical everywhere and are trivially distinguished from true mode crossings and osculations. 

\begin{rmk}
As long as we consider polynomial matrix flows, it is easy to show that two eigencurves can only cross at finitely many points. In practice, the crossing points are also \textit{relatively} rare; that is to say, two modes exhibit only a few such points (if any) over the frequency range of interest. For these reasons, they do not pose a significant challenge for computing the block-decomposition. In fact, when considering the exact continuous problem, finding one of the finitely many crossing points at an arbitrarily chosen wavenumber (out of infinitely many) has zero probability. When solving the problem numerically, the probability is finite but small, depending on the chosen threshold for when two solutions are to be considered identical. More general matrix flows can, in principle, exhibit infinitely many crossing points. Still, we would be able to determine the block-decomposition as long as there are finitely many crossings within the considered frequency range, and we test for the decomposability at sufficiently many points.
\end{rmk}

\subsection{Implementation}\noindent
The algorithm outlined in the previous sections is indeed straightforward to implement. A basic yet functional code is presented as a Matlab script in Listing~\ref{fig:matlab}. The code consists only of the main function, a function defining the matrix flow, a function performing the decomposition of the matrix flow at a given value of $k$, as well as a function for the computation of the eigencurves for each block. Each computation uses standard procedures and can be linked to the steps already discussed. The only non-obvious choice may be the use of the Dulmage-Mendelsohn permutation, implemented in the built-in Matlab function \emph{dmperm}, for identifying the block structure in the decomposition. This function has been found to be highly efficient and robust for our application and is included in the code example for compactness. However, more basic approaches to re-sorting the decomposed matrix can readily be devised, as discussed in more detail in \cite{Uhlig2020}. What is omitted for conciseness in the presented code is the treatment of repeated eigenvalues. They require comparing the eigenvalues at each $k$-value and, in the case of repeated eigenvalues, repeating the Dulmage-Mendelsohn permutation at the following $k$-value until a decomposition with strictly distinct eigenvalue has been found. Both versions of the code (the simple one in  Listing~\ref{fig:matlab} as well as a slightly more sophisticated version accounting for repeated eigenvalues) are available for download, together with the example files to reproduce the results in this paper \cite{Gravenkamp2023}.

\begin{lstlisting}[float,numbers=none,caption={Simple Matlab code for computing eigencurves of a matrix flow using block-diagonal decomposition. The code and examples are also available for download \cite{Gravenkamp2023}.},label={fig:matlab},captionpos=b]
%% computation of eigencurves utilizing uniform block-diagonalization
% input:
% E0,  E1, E2, M: coefficient matrices of matrix flow
% ka:  k-value for first decomposition
% kb:  k-value for second decomposition
% kC:  k-values for computing eigencurves
% thB: threshold for determining block structure

%% main function
function omB=eigencurves(E0,E1,E2,M,ka,kb,kC,thB)
Ea=matrixFlow(E0,E1,E2,ka);                     % evaluate matrix flow at ka
[Phi,~]=eig(Ea,M);                              % eigenvalue decomposition
Eb=matrixFlow(E0,E1,E2,kb);                     % evaluate matrix flow at kb
[ind,nBl]=decompose(Eb,Phi,thB);                % block-decomposition of Eb
omB=solveBlocks(Phi,E0,E1,E2,M,ind,nBl,kC);     % compute eigencurves of blocks
end

%% definition of matrix flow
function E=matrixFlow(E0,E1,E2,k)
E = k^2*E0 - k*E1 + E2;                         % evaluate matrix flow
end

%% decomposition of matrix flow E using eigenvectors Phi with accuracy thB
function [ind,nBl]=decompose(E,Phi,thB)
B = Phi'*E*Phi;                                 % apply transformation
B(abs(B)/norm(B)<thB)=0;                        % neglect small values
[p,~,r,~,~,~] = dmperm(B);                      % permutation
nBl = numel(r)-1;                               % number of blocks
ind=cellfun(@(i)p(r(i):r(i+1)-1),...
    num2cell(1:nBl),'UniformOutput',false);     % store block indices
end

%% compute eigencurves for each block
function omB=solveBlocks(Phi,E0,E1,E2,M,ind,nBl,kC)
omB{nBl} = [];                                  % allocate frequencies
for i = 1:nBl                                   % loop blocks
    Phic=Phi(:,ind{i});                         % eigenvectors of current block
    Mb = Phic'*M*Phic;                          % decomposed mass matrix
    omB{i} = zeros(numel(kC),numel(ind{i}));    % allocate frequencies
    for j=1:numel(kC)                           % loop k-values
        E=matrixFlow(E0,E1,E2,kC(j));           % evaluate matrix flow at k
        Eb = Phic'*E*Phic;                      % current block
        omB{i}(j,:)=sort(sqrt(eig(Eb,Mb)));     % compute eigenvalues
    end
end
end
\end{lstlisting}

\subsection{Numerical examples}\label{sec:numex1}
\subsubsection{Homogeneous isotropic layer}
\begin{figure} \centering
    \includegraphics[width=0.35\textwidth]{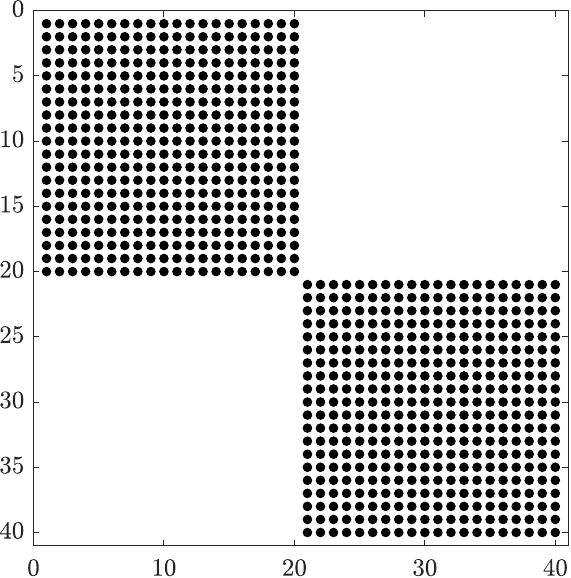}
    \caption{Block structure of the matrix $\Egv_{\mathrm{a}}^\Her \*E(k) \Egv_{\mathrm{a}}$ for the case of a homogeneous plate. \label{fig:homogeneousPlate_structure}}
\end{figure}

\begin{figure}\centering
    \includegraphics[width=0.6\textwidth]{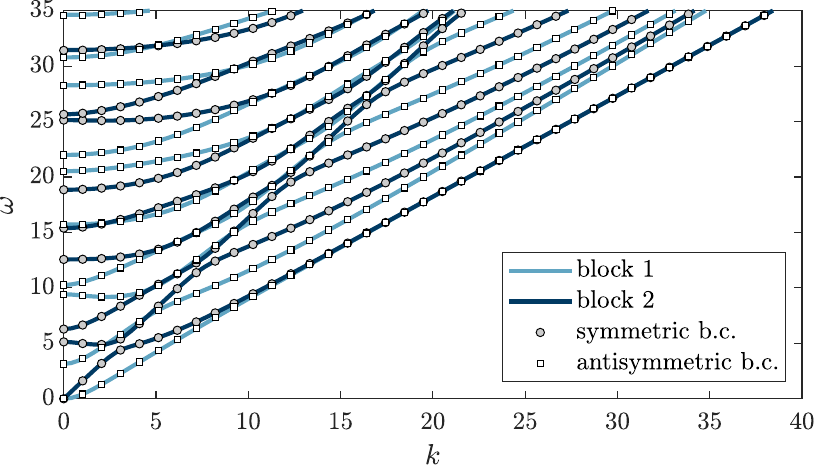}
    \caption{Dispersion curves for a homogeneous, isotropic plate. Results are computed separately for the two blocks of the block-diagonalized matrix flow. They are validated by computing the eigenvalues of the half plate with symmetric ($\circ$) and antisymmetric (\raisebox{0.5pt}{\scalebox{0.6}{$\square$}}) boundary conditions.\label{fig:homogeneousPlate_curves}}
\end{figure}

\noindent
We begin with the classical example involving a homogeneous isotropic plate with traction-free surfaces. As already mentioned in the introduction, we know that the symmetric and antisymmetric modes can exhibit various crossing points, and we expect that this behavior will also be reflected in the block structure of the associated matrix flow. We choose the material parameters: shear modulus $G=1$, mass density $\rho=1$, Poisson's ratio $\nu=0.2$ (assuming dimensionless definitions throughout). The displacement field is discretized by a single high-order element of the 'Lagrange-type' in the thickness direction (see, e.g., \cite{Gravenkamp2012} for details on this type of discretization for the same application). Here, we use an element of order 19, i.e., 20 nodes along the thickness. Following the decomposition algorithm outlined before, we perform a full eigenvalue decomposition of the matrix flow at an arbitrarily chosen value of $k_\mathrm{a}=1$ and use the computed eigenvectors for the similarity transformation at $k_\mathrm{b}=2$. The resulting block structure of the transformed matrix $\Egv_{\mathrm{a}}^\Her \*E(k) \Egv_{\mathrm{a}}$ is shown in Fig.~\ref{fig:homogeneousPlate_structure}. 
To assess the block structure, we use a threshold of $10^{-8}$, thus treating all entries with an absolute value below this threshold as zeros in the transformed matrix. We find that, for this homogeneous plate and the chosen discretization, the matrix flow can be decomposed into two blocks of equal size. Computing the eigenvalues separately for each block leads to the results shown in Fig.~\ref{fig:homogeneousPlate_curves}. As the matrices in this simple example are very small, the total computational time for obtaining the dispersion curves based on 200 $k$-values was then only about 0.04\,s on a current desktop computer (11$^\text{th}$ generation Intel i9 processors, using the Matlab implementation as published \cite{Gravenkamp2023}). In comparison, solving the complete eigenvalue problem without decomposition takes roughly twice as long.
Furthermore, we solve the same problem by discretizing only half of the plate and applying symmetric and antisymmetric boundary conditions consecutively. By comparison, we can conclude that the two blocks in the decomposition represent exactly the two sets of symmetric and antisymmetric modes.\footnote{We may note that the accuracy of both approaches in our example is different as we discretize the full and half plate using the same element order. This leads to more accurate results for the half plate (at higher computational costs as the problem is solved twice with different boundary conditions). However, the results are well converged in the presented frequency range; hence, no significant deviation is visible.}

While these results may seem expected after the previous considerations, we can make one interesting point: Not only did we discretize the full plate by only one element, but since we chose an even number of nodes, there is no node on the line of symmetry. Hence, applying (antisymmetric) boundary conditions between two nodes in the middle of an element may not have been obvious in a finite-element implementation. This can be considered an advantage of the decomposition strategy, as it allows to  `divide and conquer' the computation of dispersion curves in a very simple manner, even when the application of symmetry conditions would not be straightforward or obvious.
We can also confirm the previously mentioned result that the decomposability in this example disappears as soon as we introduce a small deviation from the model's symmetry. If we discretize the full plate by two elements but increase the shear modulus of one of the elements by 1\%, we obtain the dispersion curves already shown for illustration in the introduction in Fig.~\ref{fig:dispersionSlightlyInhomogeneous}. 
We may remark here that it could be worthwhile investigating the use of \textit{approximate} block-diago\-nali\-zation schemes, which are an active field of research with applications in signal processing, see, e.g., \cite{Maehara2011}. However, for our application, we did not, until now, find an algorithm that proved useful in providing a block-diagonalization that allows computing the eigencurves with satisfactory accuracy when the matrix flow is not exactly decomposable.

\subsubsection{Layered plate}
\begin{figure}\centering
    \subfloat[Material distribution\label{fig:compositeSketch}]{\raisebox{6.5mm}{\includegraphics[height=4cm]{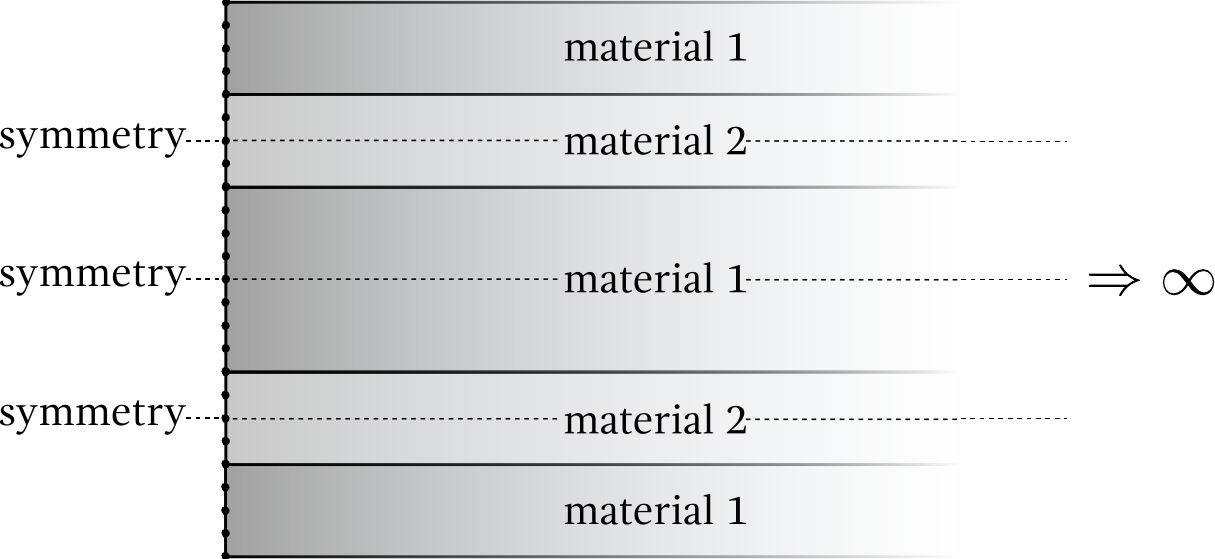}}}\qquad\qquad
    \subfloat[Block structure\label{fig:composite_block}]{\includegraphics[height=5cm]{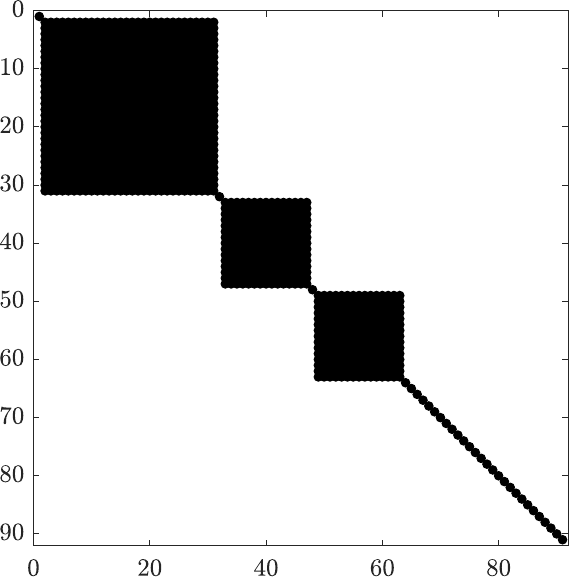}}
    \caption{Layered plate example: (a) material distribution, (b) block structure of the matrix  $\Egv_{\mathrm{a}}^\Her \*E(k) \Egv_{\mathrm{a}}$. }
\end{figure}

\begin{figure}\centering
    \subfloat[]{\includegraphics[width=0.47\textwidth]{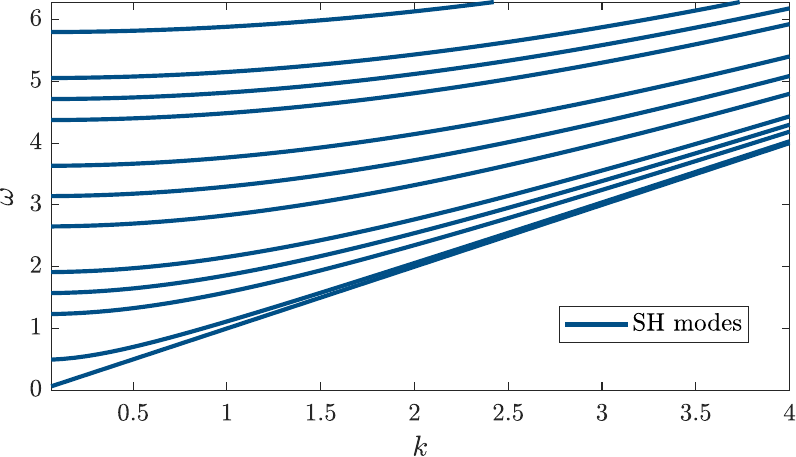}}
    \subfloat[]{\includegraphics[width=0.47\textwidth]{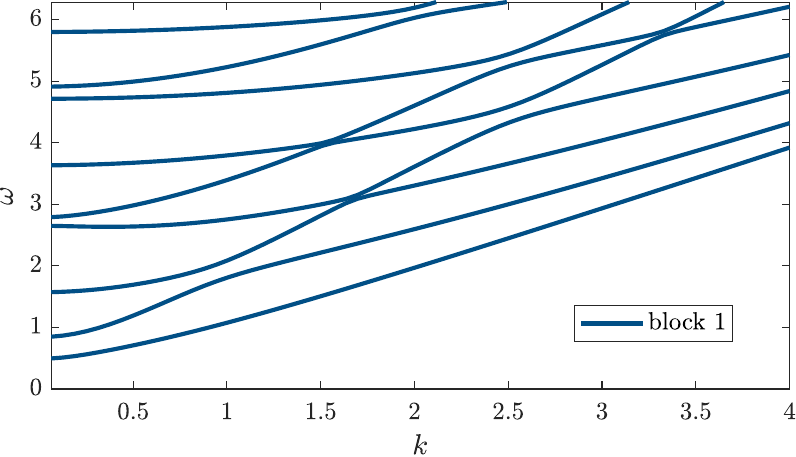}}\\
    \subfloat[]{\includegraphics[width=0.47\textwidth]{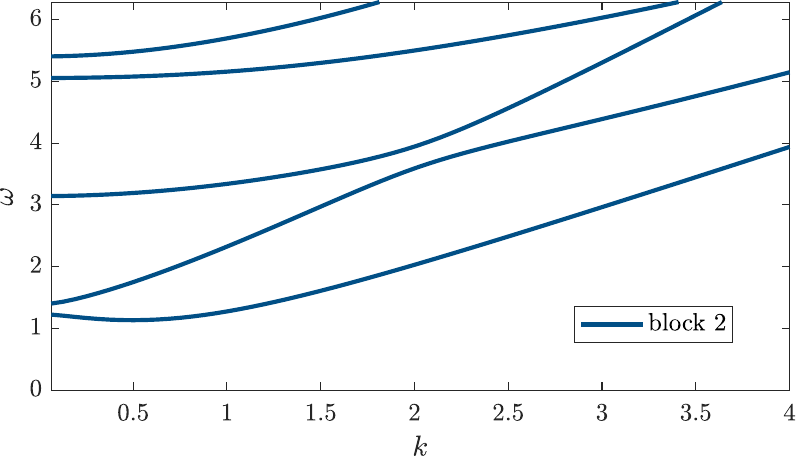}}
    \subfloat[]{\includegraphics[width=0.47\textwidth]{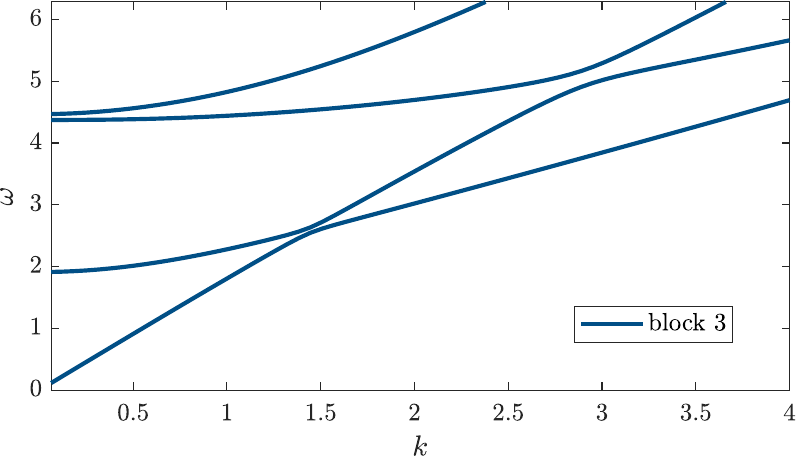}}\\
    \subfloat[]{\includegraphics[width=0.47\textwidth]{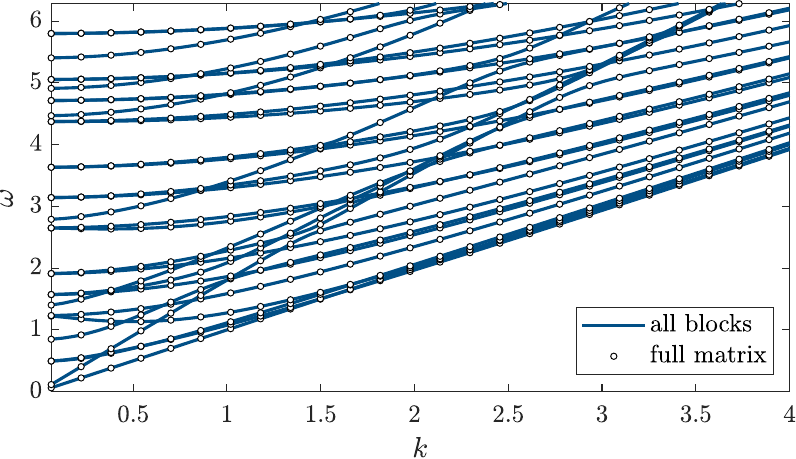}}
    \caption{Dispersion curves of a layered plate. Results are computed separately for the three blocks corresponding to the in-plane modes as well as for the shear-horizontal modes. They are validated by computing the eigenvalues of the full matrix.\label{fig:composite_curves}}
\end{figure}

\noindent
We now consider a plate consisting of two different materials in five layers as depicted in Fig.~\ref{fig:compositeSketch}. The total thickness of the layered structure is chosen as $h=6$, with the thicknesses of the individual layers (1,\,1,\,2,\,1,\,1). The structure is discretized using six line elements of a polynomial degree of 5. Vertical displacements are constrained at the top and bottom surface, and the material distribution is symmetric with respect to the plate's midplane. All three displacement components are included in the formulation; hence, we obtain not only the Lamb-type (in-plane) but also shear-horizontal (out-of-plane) modes. The material parameters are chosen as 
\begin{center}
    \begin{tabular}[h]{lccc}
        & G & $\rho$ & $\nu$\\\hline
material 1: & 1 & 1 & 0.2\\
material 2: & 2 & 2 & 0.4
\end{tabular}    
\end{center}
The resulting block structure is depicted in Fig.~\ref{fig:composite_block}. Note that the shear wave velocity $\sqrt{G/\rho}$ is the same for both materials. Thus, the shear-horizontal modes are fully decoupled and represented by blocks of unit size. The in-plane modes are represented by three blocks. As in the example of a homogeneous plate, the symmetric and antisymmetric modes are decoupled due to the line of symmetry at the plate's midplane. Furthermore, in this particular example, the block corresponding to the symmetric modes can be decomposed further into two smaller blocks as the symmetric modes exhibit another line of symmetry as indicated in Fig.~\ref{fig:compositeSketch}. This additional symmetry exists due to our choice of boundary conditions. If, for instance, we constrained the horizontal displacements at the surfaces, the block that corresponds to the antisymmetric modes could be further decomposed. The resulting dispersion diagrams are depicted in Fig.~\ref{fig:composite_curves}. We plot the eigencurves of each block and validate the results by comparing all curves against the solution of the full eigenproblem. The overall computing time using again 200 $k$-values was about 0.08\,s when employing the decomposition and 0.37\,s when solving the full eigenvalue problem at each $k$-value.

\subsubsection{Three-dimensional waveguide with symmetries}\label{numEx_3D}
\begin{figure}\centering
    \subfloat[\label{fig:squarePipe_mesh_a}]{\includegraphics[height=0.25\textwidth]{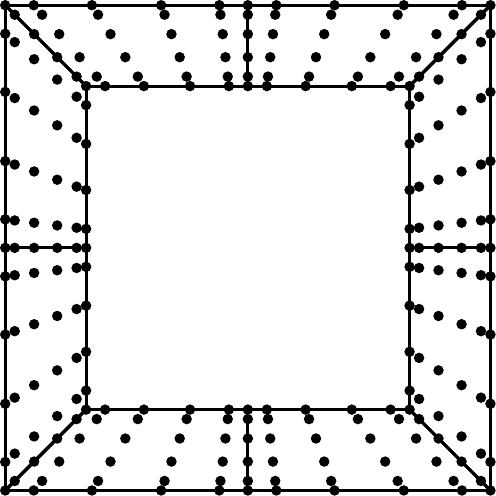}}\qquad\qquad
    \subfloat[\label{fig:squarePipe_mesh_b}]{\includegraphics[height=0.25\textwidth]{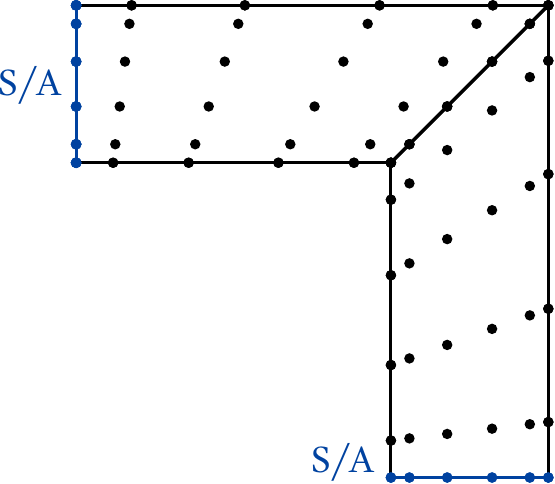}}\qquad\qquad
    \subfloat[\label{fig:squarePipe_mesh_c}]{\includegraphics[height=0.25\textwidth]{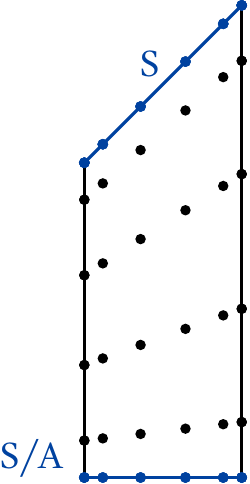}}
    \caption{Mesh representing the cross-section of a square pipe: (a) full mesh; (b) a quarter of the mesh with symmetric (S) or antisymmetric (A) boundary conditions; (c) an eighth of the mesh for computing the SS or AA modes. \label{fig:squarePipe_mesh}}
\end{figure}

\noindent
As a further example, we consider a three-dimensional structure, of which we discretize the two-dimensional cross-section, Fig.~\ref{fig:squarePipe_mesh}. The geometry represents a square pipe with a total width of 1.5 and a wall thickness of 0.25. The material parameters are $G=1$, $\rho=1$, $\nu=\nicefrac{1}{3}$.  
Similarly to the plate structures, it is worthwhile to first consider a `manual' decomposition by applying adequate boundary conditions, which is, in this case, somewhat less obvious due to the four symmetry axes at different angles.
It is naturally sufficient to reduce the mesh to a quarter of the cross-section (Fig.~\ref{fig:squarePipe_mesh_b}) and consecutively apply all combinations of symmetric and antisymmetric boundary conditions along the $x$- and $y$-axis to compute all modes. We will refer to these sets of boundary conditions as `SS', `AA', `SA', `AS' (S:~`symmetric', A: `antisymmetric'). In addition, the cases `SS' and `AA' have an extra axis of symmetry, such that we can reduce these subproblems further to an eighth of the cross-section (\ref{fig:squarePipe_mesh_c}) by again applying, in turn, symmetric and antisymmetric boundary conditions, say `SSS', `SSA', `AAS', `AAA'. Note that, in terms of implementation, the application of boundary conditions along this axis is slightly more involved, as it requires rotation of the displacement vectors on the inclined boundary in order to fix normal/tangential displacements \cite{Gravenkamp2013a}. The cases `SA' and `AS' cannot be reduced further; thus, by these symmetry arguments, we can divide the task of computing the eigenmodes into six subproblems. 

\begin{figure}\centering
    \includegraphics[width=0.6\textwidth]{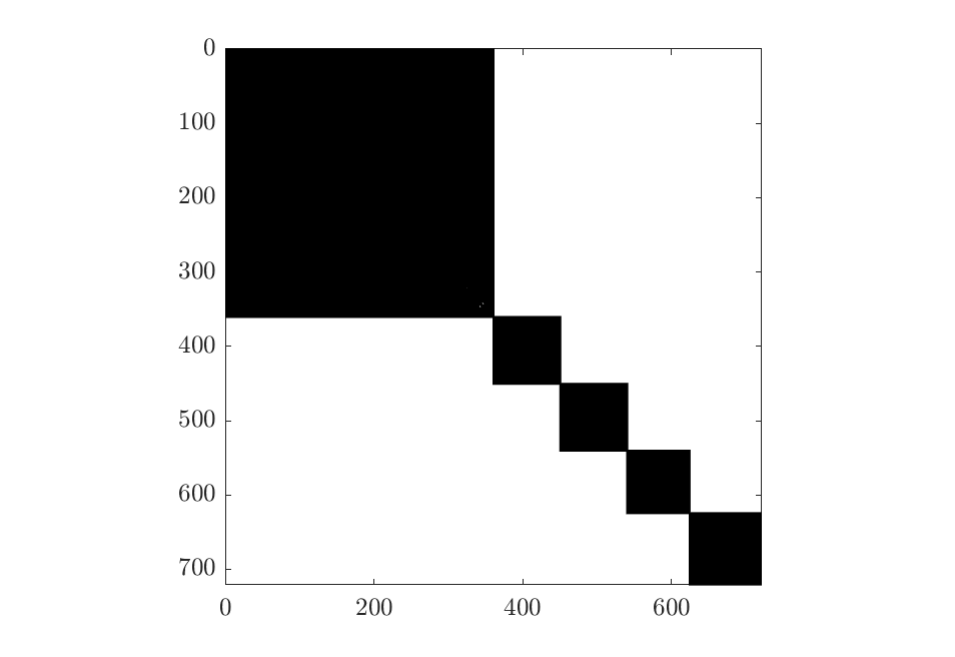}
    \caption{Block structure of the matrix $\Egv_{\mathrm{a}}^\Her \*E(k) \Egv_{\mathrm{a}}$ for the case of a square pipe. \label{fig:squarePipe_structure}}
\end{figure}

After applying the decomposition algorithm, it may come as a surprise that the resulting block-diagonal\-ized matrix contains five blocks, see Fig.~\ref{fig:squarePipe_structure}. 
The reason lies in a peculiarity of this particular example, leading to repeated eigenvalues. Due to symmetry, the eigenvalues corresponding to the sets `SA' and `AS' are \textit{identical}. Roughly speaking, we can obtain the mode shape of an `AS' mode by swapping the $x$- and $y$-components of an `SA' mode. Then, the corresponding eigenfrequencies must be the same. Hence, we obtain two sets of solutions with identical eigenvalues for any $k$ but with different eigenvectors! The large block in the decomposition corresponds to the combination of `SA' and `AS' modes and is not trivially decomposed further by the applied transformation. This result is somewhat unsatisfactory from the viewpoint of efficiency, as the bottleneck in the computation of the dispersion curves still consists in the eigenvalue decomposition of the largest block -- even though we already know that it can be subdivided by exploiting physical symmetries. However, regarding the discussion of osculations and mode crossings, this issue does not impose any additional difficulties, as the repeated eigenvalues are identical for all values of $k$.
\begin{figure}\centering
    \subfloat[]{\includegraphics[width=0.47\textwidth]{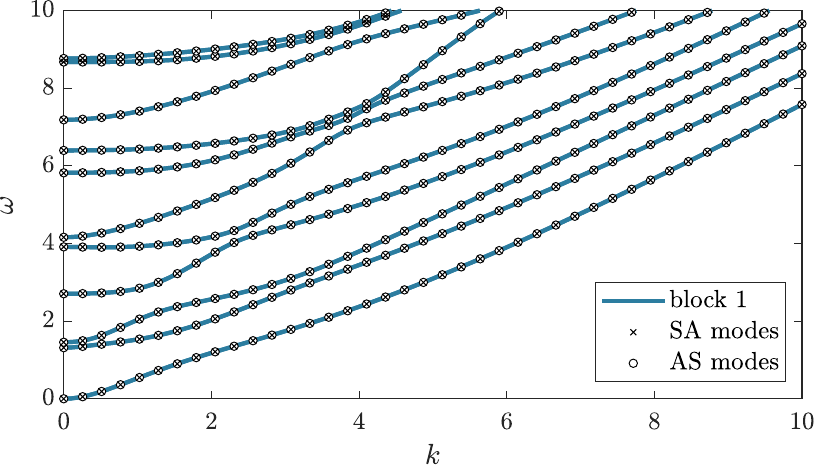}}
    \subfloat[]{\includegraphics[width=0.47\textwidth]{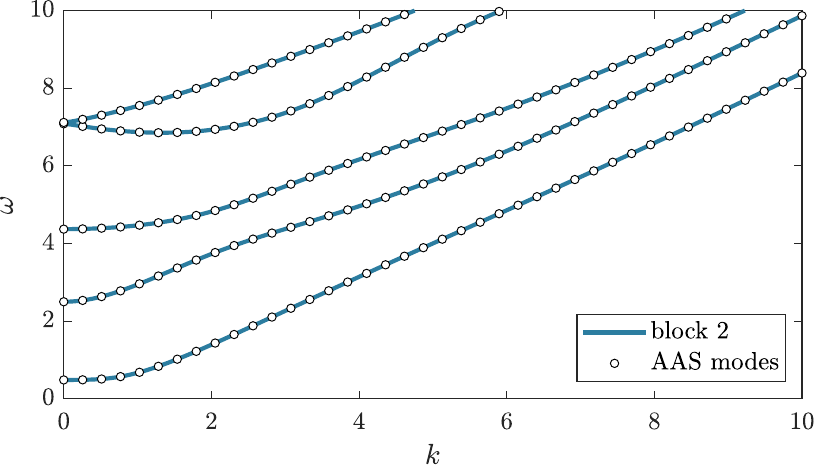}}\\
    \subfloat[]{\includegraphics[width=0.47\textwidth]{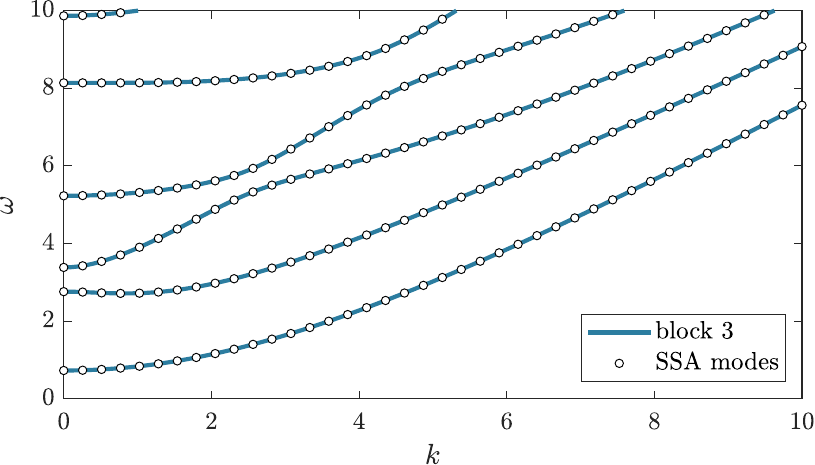}}
    \subfloat[]{\includegraphics[width=0.47\textwidth]{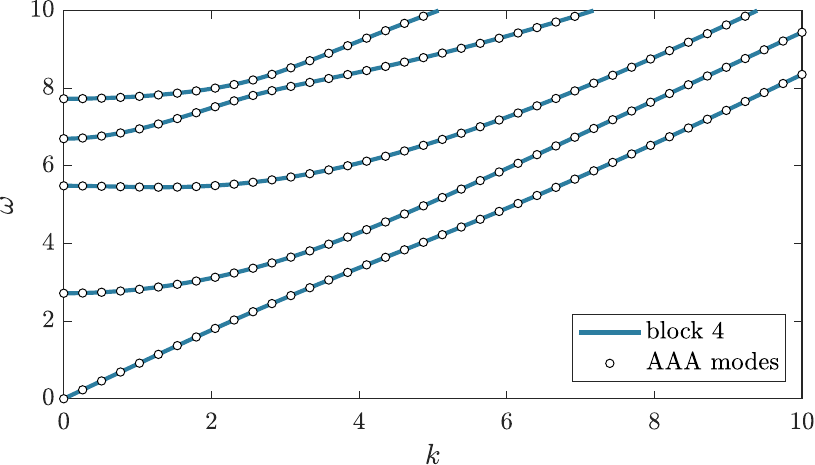}}\\
    \subfloat[]{\includegraphics[width=0.47\textwidth]{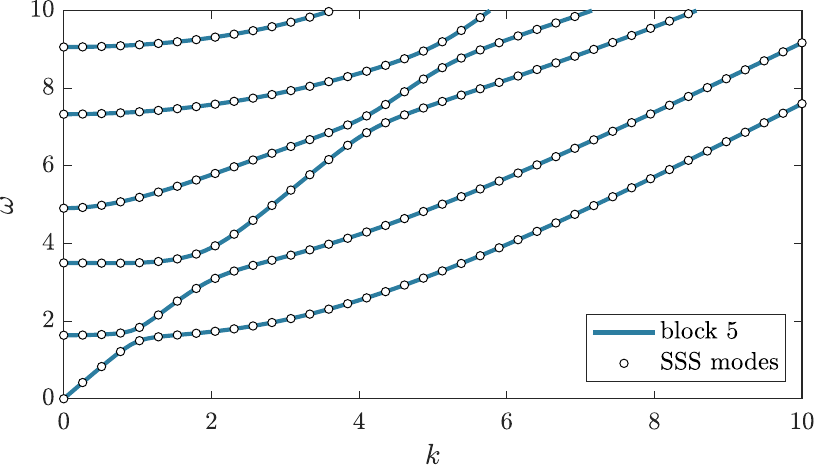}}
    \subfloat[]{\includegraphics[width=0.47\textwidth]{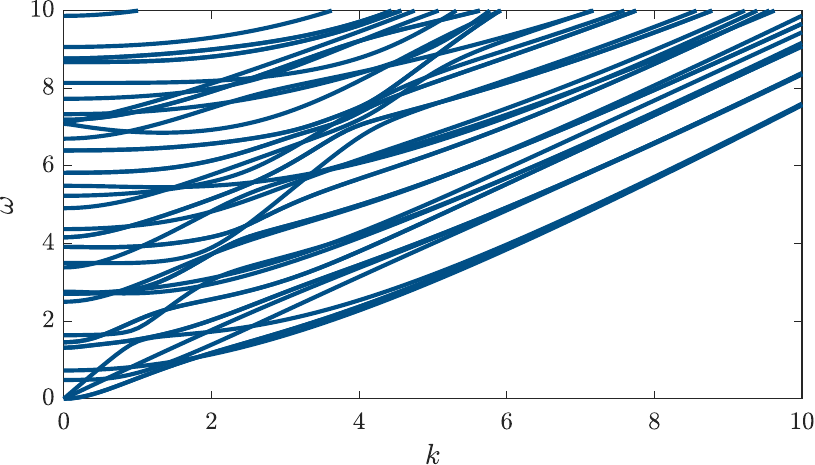}}
    \caption{Dispersion curves of a square pipe, computed individually for each block (a--e) and compared against those obtained by using symmetric and antisymmetric boundary conditions. Subfigure (f) depicts the entire mode spectrum over the selected frequency range. \label{fig:squarePipe_curves}}
\end{figure}
Figure~\ref{fig:squarePipe_curves} shows the dispersion curves obtained by computing the eigenvalues of the five blocks separately and, for comparison, by reducing the mesh and applying the different combinations of boundary conditions as discussed before. 
The eigencurves of the individual blocks match perfectly with those obtained from the different boundary conditions. We can also confirm that the solutions of the cases `SA' and `AS' are identical and match the eigencurves of the first block in the block-diagonalized matrix flow. The computational times for obtaining the dispersion curves for all blocks (again at 200 $k$-values) was about 16\,s, which is, of course, significantly larger than in the previous examples due to the relatively large matrix size of $720\times720$. Solving instead the full eigenvalue problem without prior decomposition took about ten times longer (166\,s).

\section{Computation of individual eigencurves by mode tracing}\label{sec:trace}\noindent
In this section, we apply a method for the computation of individual eigencurves that was also developed in the context of general matrix flows \cite{Uhlig2019}. The approach follows an eigencurve and computes consecutive solutions starting from an initial value at some $k$. This is in contrast to those mode tracing algorithms mentioned in the introduction that are employed in a post-processing step solely to sort the already computed solutions into sets of modes. Here, we will attempt to use the combined mode tracing and computation after assessing the block structure of the matrix flow. This facilitates the computation as the eigencurves of each block do not cross. On the other hand, we note that mode tracing can generally be difficult in the vicinity of osculations where solutions belonging to different branches can be arbitrarily similar. This potential pitfall is a drawback shared with many other approaches that use the general concept of mode tracing. For instance, in \cite{Gravenkamp2014a}, modes are computed by extrapolating previous solutions through a Pad\'e expansion and converging the same by means of inverse iteration. 
Also, in `analytical'\footnote{These methods are analytical in the sense that an exact dispersion relation is derived. However, the resulting transcendental equation needs to be solved by numerical root-finding algorithms.} approaches like the Global Matrix Method that typically involve the numerical minimization of a function, mode tracing is utilized in a similar manner to facilitate the root-finding procedure \cite{Lowe1995c,Pavlakovic1997}. It is worth remarking that, in contrast to the mentioned `analytical' methods, the following mode tracing procedure can be applied to waveguides of arbitrary cross-section and material inhomogeneity.

However, the method described in the following uses a rather different and maybe surprising formulation. 
The idea is remarkably simple and is already concisely presented in \cite{Uhlig2019} for general matrix flows. Thus, we will not repeat it in length but rather apply it to our problem at hand (again, with the trivial extension to generalized eigenvalue problems).
The method considers the residual $\*R(k)$ of the eigenvalue problem 
\begin{equation}\label{eq:residual}
    \*R(k) = \*E(k) \egv - \bar{\omega}\*M \egv
\end{equation}
for one mode and postulates the existence of a scheme that results in the residual decaying exponentially when starting from a perturbed solution. Exponential decay of $\*R(k)$ implies that
\begin{equation}\label{eq:residualDecay}
    \*R'(k) = -\eta \*R(k) 
\end{equation}
with some algorithmic constant $\eta>0$. Here and in the following, the `prime' symbol denotes a derivative with respect to $k$.
Assuming the matrix flow to be differentiable, we can evaluate the derivative $\*R'(k)$ by applying the product rule as
\begin{equation}\label{eq:residual_dk}
    \*R'(k) = \*E'(k) \egv(k) + \*E(k) \egv'(k) - \bar{\omega}'(k)\*M \egv(k) - \bar{\omega}(k)\*M \egv'(k)\,.
\end{equation}
Substituting this result into \eqref{eq:residualDecay} and re-arranging yields
\begin{equation}\label{eq:residual_ode}
    \big(\*E(k) - \bar{\omega}(k)\*M \big) \egv'(k) - \*M \egv(k)\bar{\omega}'(k)  =  -\eta \big(\*E(k) - \bar{\omega}(k)\*M\big) \egv(k) -\*E'(k) \egv(k) \,.
\end{equation}
Since the eigenvectors are determined only up to a multiplicative constant scalar, we require one additional equation to determine all components of the eigenvalues. As such an additional equation, we can choose a normalization (such as the one in Eq.~\eqref{eq:normalization_b}) and analogously assume an exponential decay of the residual in the normalization as
\begin{equation}\label{eq:residual_norm}
    2\egv^\Her(k) \*M \egv'(k) = -\mu ( \egv^\Her(k) \*M \egv(k) - 1)\,.
\end{equation}
Combining Eqs.~\eqref{eq:residual_ode} and \eqref{eq:residual_norm} yields the following system of nonlinear first-order ordinary differential equations
\begin{equation}\label{eq:residual_ode_matrix}
    \begin{bmatrix}
        \*E(k) - \bar{\omega}(k)\*M     &  -\*M \egv(k) \\
        -\egv^\Her(k) \*M               &       0
    \end{bmatrix}
    \begin{pmatrix}
        \egv'(k) \\
        \bar{\omega}'(k)
    \end{pmatrix}
     =  
     \begin{pmatrix}
        -\eta( \*E(k) - \bar{\omega}(k)\*M)\egv  -\*E'(k) \egv \\
        \mu/2 ( \egv^\Her(k) \*M \egv(k) - 1)
     \end{pmatrix}\ ,
\end{equation}
where Eq.~\eqref{eq:residual_norm} has been divided by $\shortm2$ such that the matrix in Eq.~\eqref{eq:residual_ode_matrix} is Hermitian. Equation \eqref{eq:residual_ode_matrix} holds for generalized eigenvalue problems of arbitrary matrix flows, as long as $\*M$ is independent of $k$. 
In our case, we substitute
\begin{subequations}
    \begin{align}
        \*E(k)  & =  k^2\,\*E_0 - k\,\*E_1 + \*E_2 \\
        \*E'(k) & = 2k\,\*E_0 -   \*E_1. 
    \end{align}
\end{subequations}
The resulting differential equation can be solved as an initial value problem starting from an eigen\-value\,/\,ei\-gen\-vector pair computed at some value of $k$. Countless numerical methods can be employed for the solution of this differential equation. Yang et al.~\cite{Yang2020} initially used a  variable-step solver specifically designed for stiff differential equations, implemented in Matlab's function \emph{ode15s} (see \cite{Shampine1997} for details). Uhlig later developed finite-difference schemes for this purpose \cite{Uhlig2019}. Here, we follow the first recommendation and employ the readily available \emph{ode15s} solver, which has shown to be robust for our application.

\begin{figure}\centering
    \includegraphics[width=0.6\textwidth]{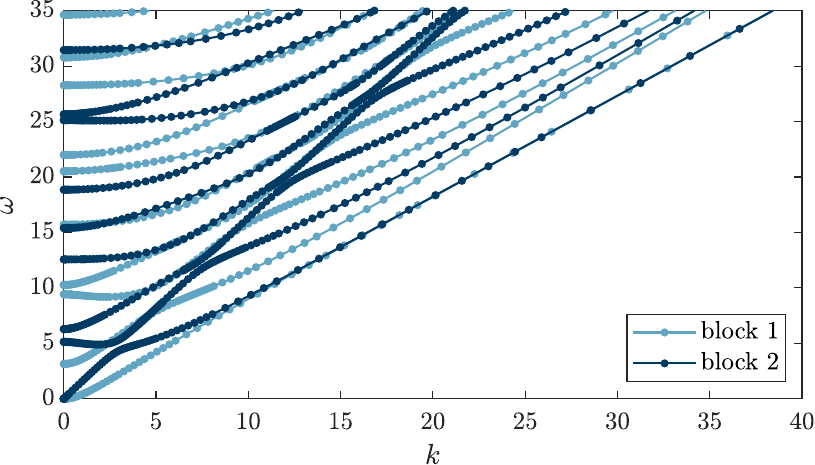}
    \caption{Dispersion curves of a homogeneous, isotropic plate, computed using the mode tracing algorithm discussed in Section~\ref{sec:trace}. \label{fig:homogeneousPlate_curves_ode}}
\end{figure}
\subsection{Numerical examples}
\begin{figure}\centering
    \subfloat[\label{fig:layeredPlate_curves_all}]{\includegraphics[height=0.36\textwidth]{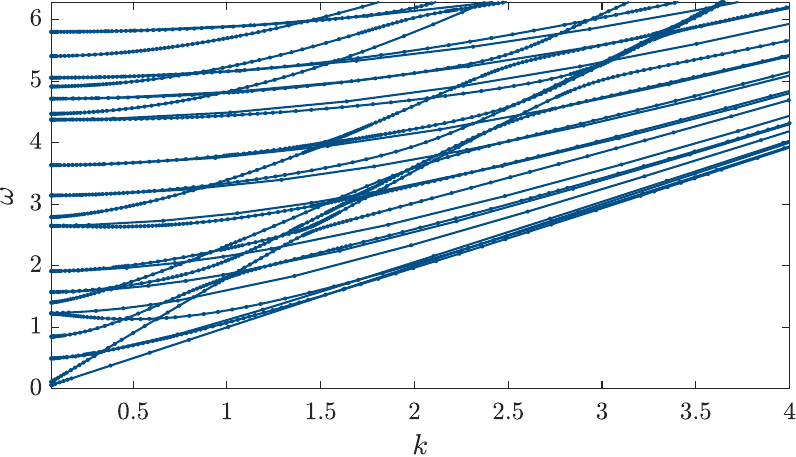}}\hfill
    \subfloat[\label{fig:layeredPlate_curves_zoom}]{\includegraphics[height=0.36\textwidth]{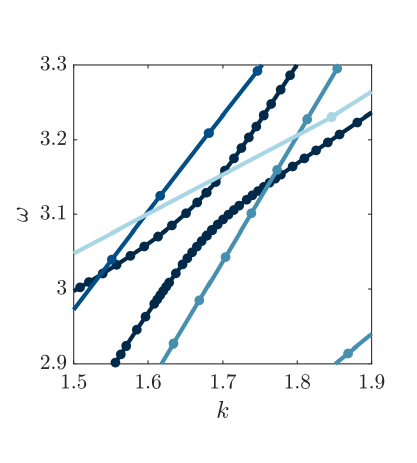}}
    \caption{(a) Dispersion curves of a layered plate, computed using the mode tracing algorithm discussed in Section~\ref{sec:trace}; (b) detail of the dispersion curves showing an osculation as well as several mode-crossings.   \label{fig:layeredPlate_curves_ode}}
\end{figure}

\begin{figure}\centering
    \includegraphics[width=0.7\textwidth]{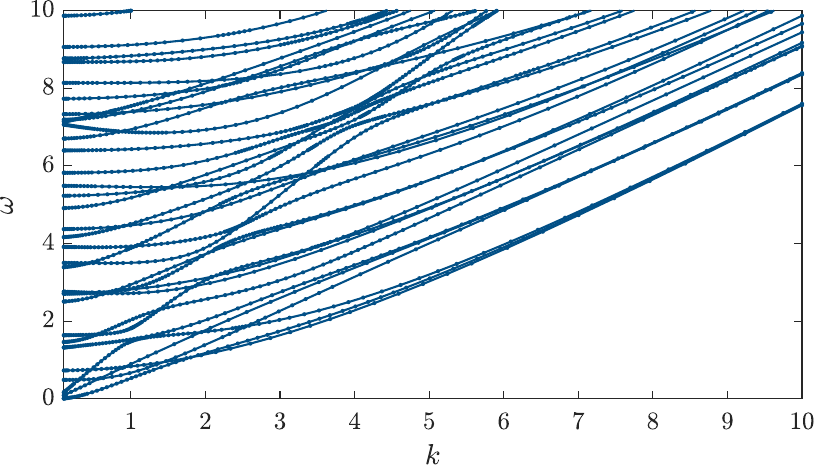}
    \caption{Dispersion curves of square pipe, computed using the mode tracing algorithm discussed in Section~\ref{sec:trace}. \label{fig:pipe_curves_ode}}
\end{figure}

\noindent
We use the same structures as in Section~\ref{sec:numex1} to demonstrate the applicability of the mode tracing algorithm. We first perform the block-decomposition discussed in Section~\ref{sec:decompose} and compute the modes for each block separately. The algorithmic parameters are chosen as $\eta=\mu = 0.001$. 
Figure~\ref{fig:homogeneousPlate_curves_ode} presents the dispersion curves of the homogeneous plate. The results agree with those obtained by the direct solution of the eigenvalue problem. The variable step size, as determined during the solution, can be seen in the figure. The step size is automatically increased for large values of $k$ where the frequencies depend almost linearly on the wavenumber.
Figure~\ref{fig:layeredPlate_curves_ode} depicts the dispersion curves of the layered plate computed in the same way, demonstrating that, despite the significantly higher complexity, the modes can be traced without additional difficulty. As an example, we show in Fig.~\ref{fig:layeredPlate_curves_zoom} an enlarged detail of \ref{fig:layeredPlate_curves_all} containing an osculation as well as several crossing modes in close proximity. 
Finally, Fig.~\ref{fig:pipe_curves_ode} shows the dispersion curves of the square pipe (Section~\ref{numEx_3D}) computed using the mode tracing algorithm. However, for solving this problem, we did not employ the matrix decomposition algorithm but rather applied all combinations of symmetric/antisymmetric boundary conditions consecutively before solving the individual problems by means of the mode tracing procedure. This has been done since the mode tracing does, so far, not work reliably for the block representing SA\,/\,AS modes, i.e., for matrices with repeated eigencurves. This is currently a shortcoming of this algorithm and a typical issue when applying mode tracing strategies to problems involving repeated eigenvalues. 

The computational times in comparison to the approach discussed in Section~\ref{sec:decompose} as well as the direct solution of the full eigenvalue problem without decomposition are summarized in Table~\ref{tab:cpuTimes}. For the method based on mode tracing, these CPU times include the computation of all modes shown in the respective figures. In general, the computing times are larger when using the mode tracing approach compared to the eigenvalue decomposition after block-diagonalization. However, mode tracing could be interesting for large problems where few modes are of relevance.
\begin{table}\centering
    \caption{\label{tab:cpuTimes}Overview of computational times for the three considered structures employing the two different approaches discussed in Sections~\ref{sec:decompose} and \ref{sec:trace}. }
    \begin{tabular}[htb]{l|c|c|c}
        & homogeneous plate & layered plate & square pipe\\
    direct solution & 0.08\,s & 0.37\,s & 166\,s\\
    decomposition (Sec.~\ref{sec:decompose}) & 0.04\,s & 0.08\,s & 16\,s\\
    mode tracing (Sec.~\ref{sec:trace}) &  0.9\,s  & 1.9\,s & 23\,s
    \end{tabular}
\end{table}

\section{Conclusion}\noindent
We have discussed and demonstrated by numerical examples that the decomposability of the matrix flow describing guided wave propagation in semi-analytical models can be tested numerically by a relatively simple algorithm. If such a decomposition exists, it can be performed straightforwardly, thus subdividing the original matrix flow into several indecomposable ones of smaller dimensions. This procedure leads, for decomposable matrix flows, to a more efficient algorithm for computing the eigencurves compared to the eigenvalue decomposition of the full matrix. In addition, this method eliminates the need for tracing or sorting modes as the eigencurves of the indecomposable subproblems do not cross. As a consequence, we can trivially distinguish (within the accuracy of our numerical model) between mode crossings and osculations. In addition, we demonstrated that the alternative approach for computing the dispersion curves by expressing them as a first-order ordinary differential equation can be used to compute individual modes and yields robust solutions.

\section{Acknowledgements}\noindent
Bor Plestenjak has been supported by the Slovenian Research Agency (grant N1-0154). Daniel A. Kiefer has been supported by LABEX WIFI (Laboratory of Excellence within the French Program ``Investments for the Future'') under Reference Nos.\ ANR-10-LABX-24 and ANR-10-IDEX-0001-02 PSL.

\appendix
\section{Characterizing the symmetry of guided waves}
\label{sec:symmetry_of_guided_waves}\noindent
The level of symmetry of waves guided in a plate can be characterized by the energy contained in the wave field's even and odd parts. To this end, consider the through-thickness displacement distributions given by $\*u(y)$, which shall be a vector containing the $x$-, $y$-, and $z$-components as required. Furthermore, assume the $y$-axis to be centered at the plate's midplane. As a first step, we decompose $\*u$ in its even and odd parts: 
\begin{subequations}
\begin{align}
    \*u_\mathrm{e}(y) &= \frac{1}{2} [ \*u(y) + \*u^*(-y) ] \,, \\
    \*u_\mathrm{o}(y) &= \frac{1}{2} [ \*u(y) - \*u^*(-y) ] \,,
\end{align}
\end{subequations}
where $(\cdot)^*$ denotes complex conjugation. Assume $\*u$ to be normalized to unit energy (in the signal-processing sense), which shall be characterized by the standard scalar product with itself, i.e.,
\begin{equation}
    \langle \*u \vert \*u \rangle = \int \*u^* \cdot \*u \,\mathrm{d} y = 1 \,.
\end{equation}
Noting that the even and odd parts are orthogonal, i.e., $\langle \*u_\mathrm{e} \vert \*u_\mathrm{o} \rangle = 0$, we have
\begin{equation}
    \langle \*u \vert \*u \rangle = \langle \*u_\mathrm{e} + \*u_\mathrm{o} \vert \*u_\mathrm{e} + \*u_\mathrm{o} \rangle = \langle \*u_\mathrm{e} \vert \*u_\mathrm{e} \rangle  + \langle \*u_\mathrm{o} \vert \*u_\mathrm{o} \rangle = 1 \,.
\end{equation}
Hence, the total energy is given by the contributions of the even and the odd parts. The latter represent precisely the contributions by \emph{symmetric} and \emph{antisymmetric} waves, respectively. Note that $\langle \*u_\mathrm{e} \vert \*u_\mathrm{e} \rangle$ and $\langle \*u_\mathrm{o} \vert \*u_\mathrm{o} \rangle$ attain values between 0 and 1. We prefer to define the \emph{level of symmetry} as
\begin{equation}
    S := \langle \*u_\mathrm{e} \vert \*u_\mathrm{e} \rangle  - \langle \*u_\mathrm{o} \vert \*u_\mathrm{o} \rangle = 2 \langle \*u_\mathrm{e} \vert \*u_\mathrm{e} \rangle -1 \,,
\end{equation}
which varies smoothly from -1 to +1 when transitioning from purely antisymmetric to purely symmetric waves.

\bibliographystyle{elsarticle-num}
\bibliography{osculation.bib, referencesDaniel.bib, referencesBor.bib}

\begin{thebibliography}{10}
\expandafter\ifx\csname url\endcsname\relax
  \def\url#1{\texttt{#1}}\fi
\expandafter\ifx\csname urlprefix\endcsname\relax\def\urlprefix{URL }\fi
\expandafter\ifx\csname href\endcsname\relax
  \def\href#1#2{#2} \def\path#1{#1}\fi

\bibitem{Mace2012}
B.~R. Mace, E.~Manconi, {Wave motion and dispersion phenomena: veering, locking
  and strong coupling effects}, The Journal of the Acoustical Society of
  America 131~(2) (2012) 1015--1028.
\newblock \href {http://dx.doi.org/10.1121/1.3672647}
  {\path{doi:10.1121/1.3672647}}.

\bibitem{Kausel2015}
E.~Kausel, P.~Malischewsky, J.~Barbosa, {Osculations of spectral lines in a
  layered medium}, Wave Motion 56 (2015) 22--42.
\newblock \href {http://dx.doi.org/10.1016/j.wavemoti.2015.01.004}
  {\path{doi:10.1016/j.wavemoti.2015.01.004}}.

\bibitem{veres_crossing_2014}
I.~A. Veres, T.~Berer, C.~Gr\"{u}nsteidl, P.~Burgholzer, On the crossing points
  of the {Lamb} modes and the maxima and minima of displacements observed at
  the surface, Ultrasonics 54~(3) (2014) 759--762.
\newblock \href {http://dx.doi.org/10.1016/j.ultras.2013.10.018}
  {\path{doi:10.1016/j.ultras.2013.10.018}}.

\bibitem{zhu_crossing_1993}
Q.~Zhu, W.~G. Mayer, On the crossing points of {Lamb} wave velocity dispersion
  curves, The Journal of the Acoustical Society of America 93~(4) (1993)
  1893--1895.
\newblock \href {http://dx.doi.org/10.1121/1.406704}
  {\path{doi:10.1121/1.406704}}.

\bibitem{kiefer_elastodynamic_2022}
D.~A. Kiefer, Elastodynamic quasi-guided waves for transit-time ultrasonic flow
  metering, FAU University Press, 2022.
\newblock \href {http://dx.doi.org/10.25593/978-3-96147-550-6}
  {\path{doi:10.25593/978-3-96147-550-6}}.

\bibitem{hernando_quintanilla_symmetry_2017}
F.~Hernando~Quintanilla, M.~J.~S. Lowe, R.~V. Craster, The symmetry and
  coupling properties of solutions in general anisotropic multilayer
  waveguides, The Journal of the Acoustical Society of America 141~(1) (2017)
  406--418.
\newblock \href {http://dx.doi.org/10.1121/1.4973543}
  {\path{doi:10.1121/1.4973543}}.

\bibitem{Uhlig2020}
F.~Uhlig, {Coalescing eigenvalues and crossing eigencurves of 1-parameter
  matrix flows}, SIAM Journal on Matrix Analysis and Applications 41~(4) (2020)
  1528--1545.
\newblock \href {http://arxiv.org/abs/2002.01274} {\path{arXiv:2002.01274}},
  \href {http://dx.doi.org/10.1137/19M1286141} {\path{doi:10.1137/19M1286141}}.

\bibitem{Uhlig2019}
F.~Uhlig, Y.~Zhang, {Time-varying matrix eigenanalyses via Zhang Neural
  Networks and look-ahead finite difference equations}, Linear Algebra and Its
  Applications 580 (2019) 417--435.
\newblock \href {http://dx.doi.org/10.1016/j.laa.2019.06.028}
  {\path{doi:10.1016/j.laa.2019.06.028}}.

\bibitem{Neumann1930}
J.~von Neumann, E.~Wigner, {On the behaviour of eigenvalues in adiabatic
  processes}, Physikalische Zeitschrift 29 (1930) 467--470.
\newblock \href {http://dx.doi.org/10.1142/9789812795762_0002}
  {\path{doi:10.1142/9789812795762_0002}}.

\bibitem{Allemang2003}
R.~J. Allemang, {The modal assurance criterion - twenty years of use and
  abuse}, Sound and Vibration 8 (2003) 14--21.

\bibitem{Gravenkamp2013a}
H.~Gravenkamp, H.~Man, C.~Song, J.~Prager, {The computation of dispersion
  relations for three-dimensional elastic waveguides using the Scaled Boundary
  Finite Element Method}, Journal of Sound and Vibration 332 (2013) 3756--3771.
\newblock \href {http://dx.doi.org/10.1016/j.jsv.2013.02.007}
  {\path{doi:10.1016/j.jsv.2013.02.007}}.

\bibitem{Krome2016}
F.~Krome, H.~Gravenkamp, {Analyzing modal behavior of guided waves using high
  order eigenvalue derivatives}, Ultrasonics 71 (2016) 75--85.
\newblock \href {http://dx.doi.org/10.1016/j.ultras.2016.05.014}
  {\path{doi:10.1016/j.ultras.2016.05.014}}.

\bibitem{Uhlig2022}
F.~Uhlig, {On the unitary block-decomposability of 1-parameter matrix flows and
  static matrices}, Numerical Algorithms 89~(2) (2022) 529--549.
\newblock \href {http://dx.doi.org/10.1007/s11075-021-01124-7}
  {\path{doi:10.1007/s11075-021-01124-7}}.

\bibitem{Bartoli2006}
I.~Bartoli, A.~Marzani, F.~{Lanza di Scalea}, E.~Viola, {Modeling wave
  propagation in damped waveguides of arbitrary cross-section}, Journal of
  Sound and Vibration 295 (2006) 685--707.
\newblock \href {http://dx.doi.org/10.1016/j.jsv.2006.01.021}
  {\path{doi:10.1016/j.jsv.2006.01.021}}.

\bibitem{Mazzotti2013b}
M.~Mazzotti, I.~Bartoli, A.~Marzani, E.~Viola, {A coupled SAFE-2.5D BEM
  approach for the dispersion analysis of damped leaky guided waves in embedded
  waveguides of arbitrary cross-section.}, Ultrasonics 53~(7) (2013)
  1227--1241.
\newblock \href {http://dx.doi.org/10.1016/j.ultras.2013.03.003}
  {\path{doi:10.1016/j.ultras.2013.03.003}}.

\bibitem{Gravenkamp2014b}
H.~Gravenkamp, C.~Birk, C.~Song, {Computation of dispersion curves for embedded
  waveguides using a dashpot boundary condition}, The Journal of the Acoustical
  Society of America 135~(3) (2014) 1127--1138.

\bibitem{Gravenkamp2015}
H.~Gravenkamp, C.~Birk, J.~Van, {Modeling ultrasonic waves in elastic
  waveguides of arbitrary cross-section embedded in infinite solid medium},
  Computers {\&} Structures 149 (2015) 61--71.
\newblock \href {http://dx.doi.org/10.1016/j.compstruc.2014.11.007}
  {\path{doi:10.1016/j.compstruc.2014.11.007}}.

\bibitem{Gravenkamp2014c}
H.~Gravenkamp, C.~Birk, C.~Song, {Numerical modeling of elastic waveguides
  coupled to infinite fluid media using exact boundary conditions}, Computers
  {\&} Structures 141 (2014) 36--45.

\bibitem{Mead1973b}
D.~J. Mead, {A general theory of harmonic wave propagation in linear periodic
  systems with multiple coupling}, Journal of Sound and Vibration 27~(2) (1973)
  235--260.
\newblock \href {http://dx.doi.org/10.1016/0022-460X(73)90064-3}
  {\path{doi:10.1016/0022-460X(73)90064-3}}.

\bibitem{Renno2013b}
J.~M. Renno, E.~Manconi, B.~R. Mace, {A Finite Element Method for Modelling
  Waves in Laminated Structures}, Advances in Structural Engineering 16~(1)
  (2013) 61--75.
\newblock \href {http://dx.doi.org/10.1260/1369-4332.16.1.61}
  {\path{doi:10.1260/1369-4332.16.1.61}}.

\bibitem{Kausel1981a}
E.~Kausel, J.~M. Ro{\"{e}}sset, J.~M. Roesset, {Stiffness matrices for layered
  soils}, Bulletin of the Seismological Society of America 71~(6) (1981)
  1743--1761.
\newblock \href {http://dx.doi.org/10.1785/BSSA0710061743}
  {\path{doi:10.1785/BSSA0710061743}}.

\bibitem{Barbosa2012a}
J.~M. {de Oliveira Barbosa}, E.~Kausel, {The thin-layer method in a
  cross-anisotropic 3D space}, International Journal for Numerical Methods in
  Engineering 89 (2012) 537--560.
\newblock \href {http://dx.doi.org/10.1002/nme.3246}
  {\path{doi:10.1002/nme.3246}}.

\bibitem{Kausel2004}
E.~Kausel, {Accurate stresses in the thin-layer method}, International Journal
  for Numerical Methods in Engineering 61~(3) (2004) 360--379.
\newblock \href {http://dx.doi.org/10.1002/nme.1067}
  {\path{doi:10.1002/nme.1067}}.

\bibitem{Gravenkamp2012}
H.~Gravenkamp, C.~Song, J.~Prager, {A numerical approach for the computation of
  dispersion relations for plate structures using the scaled boundary finite
  element method}, Journal of Sound and Vibration 331 (2012) 2543--2557.
\newblock \href {http://dx.doi.org/10.1016/j.jsv.2012.01.029}
  {\path{doi:10.1016/j.jsv.2012.01.029}}.

\bibitem{Gravenkamp2014f}
H.~Gravenkamp, C.~Birk, C.~Song, {Simulation of elastic guided waves
  interacting with defects in arbitrarily long structures using the scaled
  boundary finite element method}, Journal of Computational Physics 295 (2015)
  438--455.
\newblock \href {http://dx.doi.org/10.1016/j.jcp.2015.04.032}
  {\path{doi:10.1016/j.jcp.2015.04.032}}.

\bibitem{Itner2020a}
D.~Itner, H.~Gravenkamp, D.~Dreiling, N.~Feldmann, B.~Henning, {Efficient
  semi-analytical simulation of elastic guided waves in cylinders subject to
  arbitrary non-symmetric loads}, Ultrasonics 114 (2021) 106389.
\newblock \href {http://dx.doi.org/10.1016/j.ultras.2021.106389}
  {\path{doi:10.1016/j.ultras.2021.106389}}.

\bibitem{Hayashi2006b}
T.~Hayashi, C.~Tamayama, M.~Murase, {Wave structure analysis of guided waves in
  a bar with an arbitrary cross-section}, Ultrasonics 44 (2006) 17--24.
\newblock \href {http://dx.doi.org/10.1016/j.ultras.2005.06.006}
  {\path{doi:10.1016/j.ultras.2005.06.006}}.

\bibitem{Hayashi2003a}
T.~Hayashi, W.-J. Song, J.~L. Rose, {Guided wave dispersion curves for a bar
  with an arbitrary cross-section, a rod and rail example}, Ultrasonics 41~(3)
  (2003) 175--183.
\newblock \href {http://dx.doi.org/10.1016/S0041-624X(03)00097-0}
  {\path{doi:10.1016/S0041-624X(03)00097-0}}.

\bibitem{Thakare2017}
D.~R. Thakare, A.~Abid, D.~Pereira, J.~Fernandes, P.~Belanger, P.~Rajagopal,
  {Semi-analytical finite-element modeling approach for guided wave assessment
  of mechanical degradation in bones}, International Biomechanics 4~(1) (2017)
  17--27.
\newblock \href {http://dx.doi.org/10.1080/23335432.2017.1319295}
  {\path{doi:10.1080/23335432.2017.1319295}}.

\bibitem{Wolf1994}
J.~P. Wolf, C.~Song, {Dynamic-stiffness matrix in time domain of unbounded
  medium by infinitesimal finite element cell method}, Earthquake Engineering
  {\&} Structural Dynamics 23 (1994) 1181--1198.
\newblock \href {http://dx.doi.org/10.1002/eqe.4290231103}
  {\path{doi:10.1002/eqe.4290231103}}.

\bibitem{Song1997}
C.~Song, J.~P. Wolf, {The scaled boundary finite-element method --- alias
  consistent infinitesimal finite-element cell method --- for elastodynamics},
  Computer Methods in Applied Mechanics and Engineering 147 (1997) 329--355.
\newblock \href {http://dx.doi.org/10.1016/S0045-7825(97)00021-2}
  {\path{doi:10.1016/S0045-7825(97)00021-2}}.

\bibitem{Gravenkamp2018a}
H.~Gravenkamp, S.~Natarajan, {Scaled boundary polygons for linear
  elastodynamics}, Computer Methods in Applied Mechanics and Engineering 333
  (2018) 238--256.
\newblock \href {http://dx.doi.org/10.1016/j.cma.2018.01.031}
  {\path{doi:10.1016/j.cma.2018.01.031}}.

\bibitem{Gavric1994}
L.~Gavri{\'{c}}, {Finite element computation of dispersion properties of
  thin-walled waveguides}, Journal of Sound and Vibration 173~(1) (1994)
  113--124.
\newblock \href {http://dx.doi.org/10.1006/jsvi.1994.1221}
  {\path{doi:10.1006/jsvi.1994.1221}}.

\bibitem{Gravenkamp2019}
H.~Gravenkamp, A.~Saputra, S.~Duczek, {High-order shape functions in the scaled
  boundary finite element method revisited}, Archives of Computational Methods
  in Engineering 28 (2021) 473--494.
\newblock \href {http://dx.doi.org/0.1007/s11831-019-09385-1}
  {\path{doi:0.1007/s11831-019-09385-1}}.

\bibitem{Manconi2013b}
E.~Manconi, S.~Sorokin, {On the effect of damping on dispersion curves in
  plates}, International Journal of Solids and Structures 50~(11-12) (2013)
  1966--1973.
\newblock \href {http://dx.doi.org/10.1016/j.ijsolstr.2013.02.016}
  {\path{doi:10.1016/j.ijsolstr.2013.02.016}}.

\bibitem{Adamou2004}
A.~T.~I. Adamou, R.~V. Craster, {Spectral methods for modelling guided waves in
  elastic media}, The Journal of the Acoustical Society of America 116~(3)
  (2004) 1524--1535.
\newblock \href {http://dx.doi.org/10.1121/1.1777871}
  {\path{doi:10.1121/1.1777871}}.

\bibitem{Kiefer2019}
D.~A. Kiefer, M.~Ponschab, S.~J. Rupitsch, M.~Mayle, {Calculating the full
  leaky Lamb wave spectrum with exact fluid interaction}, The Journal of the
  Acoustical Society of America 145~(6) (2019) 3341--3350.
\newblock \href {http://dx.doi.org/10.1121/1.5109399}
  {\path{doi:10.1121/1.5109399}}.

\bibitem{kiefer_gew_2022}
D.~A. Kiefer, \href{https://github.com/dakiefer/GEW_dispersion_script}{{GEW}
  dispersion script [computer program]} (2022).
\newblock \href {http://dx.doi.org/10.5281/zenodo.7010603}
  {\path{doi:10.5281/zenodo.7010603}}.
\newline\urlprefix\url{https://github.com/dakiefer/GEW_dispersion_script}

\bibitem{Droz2014}
C.~Droz, J.-P. Lain{\'{e}}, M.~N. Ichchou, G.~Inqui{\'{e}}t{\'{e}}, {A reduced
  formulation for the free-wave propagation analysis in composite structures},
  Composite Structures 113 (2014) 134--144.
\newblock \href {http://dx.doi.org/10.1016/j.compstruct.2014.03.017}
  {\path{doi:10.1016/j.compstruct.2014.03.017}}.

\bibitem{Mitrou2016}
G.~Mitrou, N.~Ferguson, J.~M. Renno, {Wave transmission through two-dimensional
  structures by the hybrid FE/WFE approach}, Journal of Sound and Vibration 389
  (2017) 484--501.
\newblock \href {http://dx.doi.org/10.1016/j.jsv.2016.09.032}
  {\path{doi:10.1016/j.jsv.2016.09.032}}.

\bibitem{Zhu2023}
B.~Zhu, L.~Nechak, O.~Bareille, {Kriging metamodeling approach for predicting
  the dispersion curves for wave propagating in complex waveguide}, Journal of
  Sound and Vibration 551 (2023) 117595.
\newblock \href {http://dx.doi.org/10.1016/j.jsv.2023.117595}
  {\path{doi:10.1016/j.jsv.2023.117595}}.

\bibitem{Kausel1977}
E.~Kausel, J.~M. Ro{\"{e}}sset, {Semianalytic hyperelement for layered strata},
  Journal of the Engineering Mechanics Division 103 (1977) 569--588.
\newblock \href {http://dx.doi.org/10.1061/jmcea3.0002251}
  {\path{doi:10.1061/jmcea3.0002251}}.

\bibitem{Gravenkamp2016a}
H.~Gravenkamp, {A remark on the computation of shear-horizontal and torsional
  modes in elastic waveguides}, Ultrasonics 69 (2016) 25--28.
\newblock \href {http://dx.doi.org/10.1016/j.ultras.2016.03.003}
  {\path{doi:10.1016/j.ultras.2016.03.003}}.

\bibitem{Gravenkamp2023}
H.~Gravenkamp,
  \href{https://github.com/haukegravenkamp/osculations}{{Osculations [computer
  program]}} (2023).
\newblock \href {http://dx.doi.org/10.5281/zenodo.7615441}
  {\path{doi:10.5281/zenodo.7615441}}.
\newline\urlprefix\url{https://github.com/haukegravenkamp/osculations}

\bibitem{Maehara2011}
T.~Maehara, K.~Murota, {Algorithm for error-controlled simultaneous
  block-diagonalization of matrices}, SIAM Journal on Matrix Analysis and
  Applications 32~(2) (2011) 605--620.
\newblock \href {http://dx.doi.org/10.1137/090779966}
  {\path{doi:10.1137/090779966}}.

\bibitem{Gravenkamp2014a}
H.~Gravenkamp, F.~Bause, C.~Song, {On the computation of dispersion curves for
  axisymmetric elastic waveguides using the Scaled Boundary finite Element
  Method}, Computers {\&} Structures 131 (2014) 46--55.
\newblock \href {http://dx.doi.org/10.1016/j.compstruc.2013.10.014}
  {\path{doi:10.1016/j.compstruc.2013.10.014}}.

\bibitem{Lowe1995c}
M.~J.~S. Lowe, {Matrix techniques for modeling ultrasonic waves in multilayered
  media}, IEEE Transactions on Ultrasonics, Ferroelectrics and Frequency
  Control 42~(4) (1995) 525--542.
\newblock \href {http://dx.doi.org/10.1109/58.393096}
  {\path{doi:10.1109/58.393096}}.

\bibitem{Pavlakovic1997}
B.~Pavlakovic, M.~J.~S. Lowe, D.~N. Alleyne, {Disperse: A general purpose
  program for creating dispersion curves}, in: D.~O. Thompson, D.~E. Chimenti
  (Eds.), Review of Progress in Quantitative NDE, Plenum Press, 1997, pp.
  185--192.

\bibitem{Yang2020}
M.~Yang, C.~Li, F.~Uhlig, H.~Hu, Y.~Zhang, B.~Qiu, {Continuous ZNN Models for
  Computation of Time-Varying Eigenvalues and Corresponding Eigenvectors}, in:
  IEEE International Conference on Mechatronics and Automation, 2020, pp.
  1523--1528.
\newblock \href {http://dx.doi.org/10.1109/ICMA49215.2020.9233731}
  {\path{doi:10.1109/ICMA49215.2020.9233731}}.

\bibitem{Shampine1997}
L.~F. Shampine, M.~W. Reichelt, {The Matlab ode suite}, Siam J. Sci. Comput.
  18~(1) (1997) 1--22.
\newblock \href {http://dx.doi.org/10.1137/S1064827594276424}
  {\path{doi:10.1137/S1064827594276424}}.

\end{thebibliography}

\end{document}